\documentclass[twocolumn]{aastex631}

\usepackage{graphicx}
\usepackage{amsmath}
\usepackage{amssymb}
\usepackage{xspace}

\newcommand{\chandra}{{\it Chandra}\xspace}

\newcommand{\xmm}{{\it XMM-Newton}\xspace}

\newcommand{\nustar}{\textit{NuSTAR}\xspace}

\newcommand{\Mone}{M82\,X-1\xspace}
\newcommand{\Mtwo}{M82\,X-2\xspace}

\newcommand{\asec}{\ensuremath{^{\prime\prime}}}
\newcommand{\amin}{\ensuremath{^{\prime}}}

\newcommand{\ledd}{\ensuremath{L_{\rm Edd}}}

\newcommand{\pspin}{\ensuremath{p_{\rm spin}}}

\newcommand{\nudot}{\ensuremath{\dot{\nu}}}
\newcommand{\Porbdot}{\ensuremath{\dot{P}_{\rm orb}}}
\newcommand{\Porb}{\ensuremath{P_{\rm orb}}}
\newcommand{\Mdot}{\ensuremath{\dot{M}}}
\newcommand{\mdot}{\ensuremath{\dot{m}}}
\newcommand{\tasc}{\ensuremath{T_{\rm asc}{}}}
\newcommand{\rin}{\ensuremath{R_{\rm in}}}
\newcommand{\rco}{\ensuremath{R_{\rm co}}}
\newcommand{\rsph}{\ensuremath{R_{\rm sph}}}

\newcommand{\msun}{\ensuremath{M_{\odot}}\xspace}

\newcommand{\msix}{\ensuremath{10^{-6}}\xspace}
\newcommand{\yrmone}{\ensuremath{\mathrm{yr^{-1}}}\xspace}
\newcommand{\msunyr}{\ensuremath{\msun\,\yrmone}\xspace}
\newcommand{\msixyrmone}{\ensuremath{\msix\,\yrmone}\xspace}

\newcommand{\figref}{Figure~\ref}
\newcommand{\secref}{Section~\ref}
\newcommand{\tabref}{Table~\ref}

\shorttitle{Orbital decay in M82 X-2}
\shortauthors{Bachetti et al.}
\graphicspath{{./}{figures/}}

\begin{document}

\title{Orbital decay in M82 X-2}

\author[0000-0002-4576-9337]{Matteo Bachetti}
\email{matteo.bachetti@inaf.it}
\affiliation{INAF-Osservatorio Astronomico di Cagliari, via della Scienza 5, I-09047 Selargius (CA), Italy}

\author{Marianne Heida}
\affiliation{European Southern Observatory, Karl-Schwarzschild-Strasse 2, 85748 Garching bei München, Germany}

\author{Thomas Maccarone}
\affiliation{Department of Physics and Astronomy, Texas Tech University, Lubbock, TX, USA}

\author{Daniela Huppenkothen}
\affiliation{SRON Netherlands Institute for Space Research, Sorbonnelaan 2, 3584 CA, Utrecht, Netherlands}

\author{Gian Luca Israel}
\affiliation{INAF-Osservatorio Astronomico di Roma, via Frascati 33, I-00078 Monteporzio Catone, Italy}

\author{Didier Barret}
\affiliation{IRAP, Université de Toulouse, CNRS, CNES, 9 avenue du Colonel Roche, 31028, Toulouse, France}

\author{Murray Brightman}
\affiliation{Cahill Center for Astronomy and Astrophysics, California Institute of Technology, Pasadena, CA 91125, USA}

\author{McKinley Brumback}
\affiliation{Cahill Center for Astronomy and Astrophysics, California Institute of Technology, Pasadena, CA 91125, USA}

\author{Hannah P. Earnshaw}
\affiliation{Cahill Center for Astronomy and Astrophysics, California Institute of Technology, Pasadena, CA 91125, USA}

\author{Karl Forster}
\affiliation{Cahill Center for Astronomy and Astrophysics, California Institute of Technology, Pasadena, CA 91125, USA}

\author{Felix F\"urst}
\affiliation{Quasar Science Resources S.L for European Space Agency (ESA), ESAC, Camino Bajo del Castillo s/n, 28692 Villanueva de la Ca\~nada, Madrid, Spain}

\author{Brian W. Grefenstette}
\affiliation{Cahill Center for Astronomy and Astrophysics, California Institute of Technology, Pasadena, CA 91125, USA}

\author{Fiona A. Harrison}
\affiliation{Cahill Center for Astronomy and Astrophysics, California Institute of Technology, Pasadena, CA 91125, USA}

\author{Amruta D. Jaodand}
\affiliation{Cahill Center for Astronomy and Astrophysics, California Institute of Technology, Pasadena, CA 91125, USA}

\author{Kristin K. Madsen}
\affiliation{CRESST and X-ray Astrophysics Laboratory, NASA Goddard Space Flight Center, Greenbelt, MD 20771, USA}

\author{Matthew Middleton}
\affiliation{Department of Physics and Astronomy, University of Southampton, Highfield, Southampton SO17 1BJ, UK}

\author{Sean N. Pike}
\affiliation{Cahill Center for Astronomy and Astrophysics, California Institute of Technology, Pasadena, CA 91125, USA}

\author{Maura Pilia}
\affiliation{INAF-Osservatorio Astronomico di Cagliari, via della Scienza 5, I-09047 Selargius (CA), Italy}

\author{Juri Poutanen}
\affiliation{Department of Physics and Astronomy,  FI-20014 University of Turku, Finland}
\affiliation{Space Research Institute of the Russian Academy of Sciences, Profsoyuznaya Str. 84/32, Moscow 117997, Russia}
\affiliation{Nordita, KTH Royal Institute of Technology and Stockholm University, Roslagstullsbacken 23, SE-10691 Stockholm, Sweden}

\author{Daniel Stern}
\affiliation{Cahill Center for Astronomy and Astrophysics, California Institute of Technology, Pasadena, CA 91125, USA}

\author{John A. Tomsick}
\affiliation{Space Sciences Laboratory, University of California, 7 Gauss Way, Berkeley, CA 94720-7450, USA}

\author{Dominic J. Walton}
\affiliation{Institute of Astronomy, Madingley Road, Cambridge, CB3 0HA, UK}
\affiliation{Centre for Astrophysics Research, University of Hertfordshire, College Lane, Hatfield AL10 9AB, UK}

\author{Natalie Webb}
\affiliation{IRAP, Université de Toulouse, CNRS, CNES, 9 avenue du Colonel Roche, 31028, Toulouse, France}

\author{J\"orn Wilms}
\affiliation{Remeis-Observatory and Erlangen Centre for Astroparticle Physics, Friedrich-Alexander-Universit\"at Erlangen-N\"urnberg, Sternwartstr. 7, 96049 Bamberg}



\begin{abstract}

    \Mtwo is the first pulsating ultraluminous X-ray source (PULX) discovered.
    The luminosity of these extreme pulsars, if isotropic, implies an extreme mass transfer rate. An alternative is to assume a much lower mass transfer rate, but with an apparent luminosity boosted by geometrical beaming.
    Only an independent measurement of the mass transfer rate can help discriminate between these two scenarios.
    In this Paper, we follow the orbit of the neutron star for seven years, measure the decay of the orbit ($\Porbdot/\Porb\approx-8\cdot\msixyrmone$), and argue that this orbital decay is driven by extreme mass transfer of more than 150 times the mass transfer limit set by the Eddington luminosity.
    If this is true, the mass available to the accretor is more than enough to justify its luminosity, with no need for beaming.
    This also strongly favors models where the accretor is a highly-magnetized neutron star.
\end{abstract}



\section{Introduction} \label{sec:intro}
The luminosity of accreting sources is largely driven by the amount of matter that is transferred onto the accreting object, whether it be from a donor star for typical neutron stars and stellar mass black holes, or an accretion disk for supermassive black holes at the centers of galaxies \citep{frankAccretionPowerAstrophysics2002}.
There is a classical limit to the mass transfer, which corresponds to the mass accretion rate that leads to a balance between the force of radiation pressure pushing outward and the gravitational force acting inward on an accreting object of mass $M$.
For spherical hydrogen accretion, this corresponds to the Eddington luminosity:
\begin{equation}
    \ledd \approx 1.3\cdot10^{38} \frac{M}{\msun}\,\mathrm{erg\,s^{-1}}
\end{equation}
Therefore, the extreme luminosity of ultraluminous X-ray sources (ULXs; \citealt{kaaretUltraluminousXRaySources2017,fabrikaUltraluminousXRaySources2021}) led many to think that these sources were powered by intermediate-mass black holes.
Over the years, multiple pieces of evidence cast doubt on the applicability of this classical limit on ULXs \citep{poutanenSupercriticallyAccretingStellar2007,gladstoneUltraluminousState2009,bachettiUltraluminousXRaySources2013}.
Eventually, the discovery of pulsating ultraluminous X-ray sources (PULXs; \citealt{bachettiUltraluminousXraySource2014}, hereafter B14), accreting neutron stars radiating hundreds of times above their Eddington limits, demonstrated that super-Eddington accretion was a viable explanation for the majority of ULXs.
It is still unclear how these pulsars (pulsating neutron stars) emit this extreme luminosity.
Some argue that the isotropic luminosity is much lower, and the observed luminosity is boosted by geometrical beaming, driven by the collimation of a (less extreme) super-Eddington disk \citep{kingPulsingULXsTip2017}.
This interpretation has found some support in global MHD simulations of accreting black holes and neutron stars, where mild to extreme geometrical beaming is observed (e.g. \citealt{jiangGlobalThreedimensionalRadiation2014,abarcaBeamedEmissionNeutronstar2021}).
However, these simulations assume a low magnetic field of the neutron star ($\lesssim10^{10}$\,G), if any, and this collimation effect is likely to be lessened when the magnetic field of the pulsar is stronger.
In fact, other models explain the luminosity with arguments centered on a high magnetic field of the pulsar ($>10^{13}$\,G), like the reduction of the Thomson scattering cross section in high magnetic fields, either in their dipolar \citep{mushtukovMaximumAccretionLuminosity2015, mushtukovOpticallyThickEnvelopes2017} or their multipolar components \citep{briceSupereddingtonEmissionAccreting2021}.
This reduction of the cross section allows to hit the local Eddington limit at much higher mass accretion rates, increasing the maximum luminosity.
It is also possible that the solution is a mixture of genuine super-Eddington accretion and a small amount of beaming \citep{israelAccretingPulsarExtreme2017}.

A key difference between these models is the relation that they assume between the mass accretion rate $\mdot=\Mdot/\Mdot_{\rm Edd}$ and the luminosity, linear in the low-beaming scenario, almost quadratic ($L\propto(1 + \log \mdot) \mdot^2$) in the other, due to the assumed quadratic dependence of beaming on the mass accretion rate \citep{kingAccretionRatesBeaming2008}. In other words, beaming models infer a much lower mass transfer rate between the donor star and the neutron star for a given luminosity.

An independent measurement of the mass transfer is key for disentangling these two scenarios.
In principle, one way to measure this transfer of matter between two orbiting objects is through the observation of a decay of the orbital period \citep{taurisFormationEvolutionCompact2006}.

One of the best systems where this can be tested is \Mtwo, the first pulsating ultraluminous X-ray source ever discovered.
B14 and \citet{bachettiAllOnceTransient2020} (hereafter B20) measured the orbit of this PULX very precisely, determining an orbital period of $2.532948(4)$\,d, a semi-major axis of $22.215(5)$\,light-sec, and no detectable eccentricity ($<$0.003).
What makes this system particularly interesting from the point of view of orbital decay measurements is that its revolution period is short enough, and the ephemeris known so precisely, that the epoch of passage through the ascending node can be constrained to $\sim100$\,s with a single, reasonably long, X-ray observation.

In this Paper, by tracking the ascending node passages over 8 years of \nustar observations, we present the precise measurement of orbital decay in \Mtwo, leading to an estimate of mass transfer whose value agrees to within a factor 2 to the one inferred from the luminosity of the pulsar.

In \secref{sec:reduction} we describe the data reduction and in \secref{sec:timing} we detail the temporal analysis that led to the orbital decay measurement, while we devote the last sections to the interpretation of this orbital decay.

\section{Data reduction}\label{sec:reduction}
\subsection{NuSTAR}
We downloaded all the \nustar data of M82 from the High Energy Astrophysics Science Archive Research Center (HEASARC).
We ran \texttt{nupipeline} with standard options to produce cleaned event files.
This tool produces different event files corresponding to different observing modes: \texttt{SCIENCE} (01), \texttt{OCCULTATION} (02), \texttt{SLEW} (03), \texttt{SAA} (04), \texttt{CALIBRATION} (05), and \texttt{SCIENCE\_SC} (06).
The modes usable for science are 01 and 06.
Note, however, that only mode-01 data are recorded in normal instrumental conditions. Mode-06 data correspond to time intervals where only a subset of the camera head units (CHUs) are available, and the astrometry can be off by 1--2$\amin$ (See \citealt{waltonSoftStateCygnus2016} for an example of the astrometry issues in this observing mode).
For mode-01 data, we used a region of 70$\asec$ around the centroid of the X-ray source corresponding to the position of \Mone and \Mtwo, which is spatially unresolved in \nustar.
The centroid was calculated independently for each observation and for each of the two focal plane modules, as a mismatch of $\sim10\asec$ can be expected.

We processed mode-06 data with the \texttt{nusplitsc} tool, which separates events corresponding to different CHU combinations. For each of these event files, we adjusted the centroid of the source and repeated the selection done for mode-01 data.
Finally, we merged the source-selected event lists from mode-01 and mode-06 data.
In only a few cases, due to the source falling on a chip gap, we saw that the light curve showed visible ``steps'' between intervals corresponding to different CHU combinations.
We verified that the addition or elimination of the problematic intervals did not alter significantly the power around the pulsation frequency $\sim$0.7 Hz.

Finally, we ran \texttt{barycorr} to refer the photon arrival times to the solar system barycenter.
We selected the ICRS reference frame, the DE421 JPL ephemeris, and the position of \Mtwo determined by \chandra.
For all observations, we used the latest clock correction file available, that provides an absolute time precision of $\sim60\,\mu$s.

\section{Timing analysis}\label{sec:timing}

\begin{figure*}[ht]
    \centering
    \includegraphics[width=0.46\linewidth]{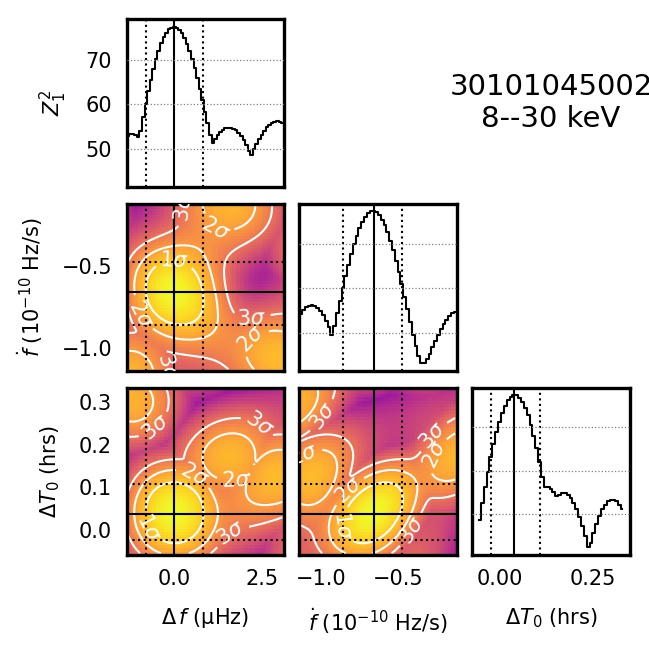}
    \includegraphics[width=0.46\linewidth]{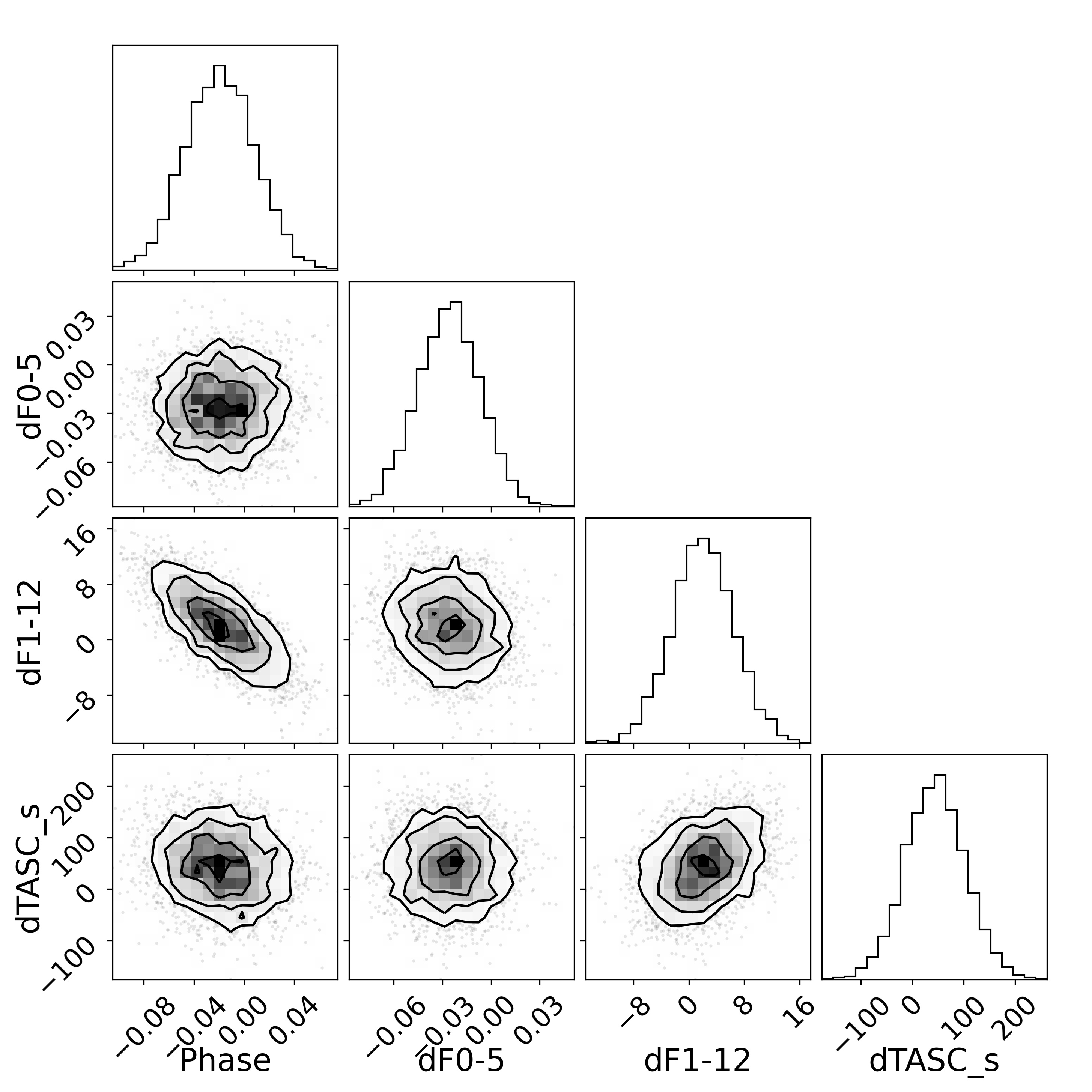}
    \caption{Example detection of pulsations, and orbital/spin parameter refinement, from ObsID 30101045002.
    The color map shows the Rayleigh search in a three-parameter grid, using spin-frequency, spin first derivative, and the drift of the periastron passage $\tasc$.
    The corner plot on the right, instead, shows the refinement of the results of the Rayleigh search, with the addition of pulse phase, fitted using the Bayesian method by \citet{pletschGammaRayTimingRedback2015}.
    E.g.: if \texttt{dF0-5} is -0.03 and the initial F0 was 0.725, this means that the best-fit frequency is $0.725 -0.03 \cdot 10^{-5}$ Hz}
    \label{fig:search}
\end{figure*}

\begin{table*}[ht]
    \centering
    \begin{tabular}{l c c c c c c c}
\hline
\hline
Obs. ID & Epoch & Energy & $T_{\rm asc}$ & $f_{\rm spin}$ & $\dot{f}_{\rm spin}$ & $\ddot{f}_{\rm spin}$ & $\Delta\,T_{\rm asc}$ (s) \\
 & MJD & keV & MJD & Hz& $10^{-12}$\,Hz\,s$^{-1}$ & $10^{-16}$\,Hz\,s$^{-2}$ & s \\
\hline
80002092002$^{*}$ & 56681.24441 &8--30 & 56682.073(5)  & 0.728509(4)  & -90(70) &  & 600(500)  \\
80002092004 & 56683.81009 &8--30 & 56684.6004(28)  & 0.7285316(16)  & 50(35)  &  & 110(240)  \\
80002092006 & 56688.80899 &8--30 & 56689.6656(5)  & 0.72854791(22)  & 15.3(12)  & 1.20(27)  & 50(40)  \\
80002092007 & 56694.12259 &3--30 & 56694.73184(19)  & 0.72856174(6)  & 34.6(14)  &  & 86(17)  \\
80002092007 & 56697.38070 &3--30 & 56697.2648(4)  & 0.72857943(8)  & 92(6)  &  & 90(40)  \\
80002092008 & 56700.75316 &3--30 & 56699.798(5)  & 0.728609(5)  & 90(60)  &  & 100(400)  \\
80002092009 & 56700.75316 &3--30 & 56699.7978(4)  & 0.72860925(13)  & 106(5)  &  & 90(32)  \\
80002092011 & 56720.87754 &8--30 & 56720.0612(12)  & 0.7287596(6)  & 113(13)  &  & 80(100)  \\
30101045002$^{*}$ & 57495.31178 &8--30 & 57495.1440(7)  & 0.72519103(20)  & \textbf{-65(5) } &  & 140(60)  \\
90201037002 & 57641.99852 &8--30 & 57642.049(8)  & 0.7239040(12)  & -320(230) &  & -400(700)  \\
30502021002$^{*}$ & 58919.09530 &3--30 & 58918.6261(12)  & 0.7219294(7)  & 36(15)  &  & -2860(100)  \\
30602027002$^{*}$ & 59312.65089 &3--30 & 59313.750(6)  & 0.7222096(21)  & 120(150)  &  & -4300(500)  \\
30602027004$^{*}$ & 59326.05680 &8--30 & 59326.409(4)  & 0.7222978(25)  & 50(70)  &  & -4700(400)  \\
30702012002$^{*}$ & 59505.27806 &3--30 & 59506.2428(11)  & 0.72086594(33)  & \textbf{-45(14) } &  & -5210(90)  \\
\hline
    \end{tabular}
    \caption{Spin and orbital parameters from the multidimensional timing procedure in \secref{sec:timing}.
    Starred ObsIDs are those corresponding to new detections from this paper.
    We also highlight in bold observations with significant ($>3\sigma$) evidence of spin down.
    The energy range is the one where the pulsations are detected with the highest significance.
    Data from ObsID 80002092006 start after the glitch reported by B20 at MJD 56685.7
    ObsID 80002092007 has a sudden change of frequency, probably another glitch, around MJD 56696, therefore we split the observation in two parts around that epoch.
    $\Delta\,T_{\rm asc}$ was calculated with respect to the orbital ephemeris from B20}
    \label{tab:obs}
\end{table*}

Due to the very low pulsed fraction in the \xmm band, demonstrated in Appendix~\ref{sec:energy}, we only used \nustar data for Timing analysis.

Initially, we largely followed the search strategies used in B20, running $Z^2_1$ searches \citep{buccheriSearchPulsedGammaray1983}, also known as the Rayleigh test, on the event arrival times corrected for orbital motion, varying the ascending node passage epoch $\tasc$ on a fine grid between $-\Porb/5$ and $\Porb/5$.
This time, the search allowed a range of spin derivatives \textit{for each trial ascending node passage value}.
The spin parameters vary so rapidly that they are only loosely constrained when observations are just a $\sim$week apart. There is no way to reliably phase connect separate observations. Therefore, even observations done $ \sim2$ weeks apart were analyzed singularly.
For the search in the $f-\dot{f}$ plane, we used the ``quasi-fast folding algorithm'' (B20), that calculates the $Z^2_1$ on pre-binned profiles \citep{bachettiExtendingStatisticsGeneric2021}, using at least 16 bins for the folded profiles.
Moreover, we ran the search both in the full energy band and between 8 and 30\,keV.
This allowed four detections in the new observations.
We also re-ran a pulsation search using all available observations, and we found highly significant ($\geq5\sigma$) pulsations in two archival datasets, corresponding to ObsIDs 30101045002 and 80002092002.
In both observations, pulsations are more strongly detected in the 8--30 keV energy band.
Moreover, surprisingly, we find that during 30101045002 the pulsar was instantaneously spinning \textit{down}.
This is the first time that \Mtwo\ is found spinning down while accreting and pulsating, and provides clear evidence that a significant part of the torque from the disk is happening outside the corotation radius (see Appendix~\ref{sec:radii} for more details).
This is probably why B20 only obtained marginal evidence for pulsations in this ObsID.
This new detection is important because pulsations are detected over a $\sim4$-day interval, which is long enough to provide an excellent constraint on $\tasc$.
For each detection, we then ran the search again around the best solution, oversampling by a large factor to find the best estimate of the mean of each parameter.

We proceeded to create local timing solutions for each observation, leaving the orbital parameters from B20 unchanged with the exception of the ascending node, and setting the current spin frequency and frequency derivative.
These local solutions differed only for the parameters F0 (spin frequency in Hz), F1 (spin derivative in Hz/s), and TASC (ascending node passage in MJD).
Then, we used the method by \citet{pletschGammaRayTimingRedback2015} to make a Bayesian fit of the local timing solution, using a sinusoidal pulse template normalized to the same pulsed fraction of the pulsar in each given observation.
Working in phase space instead of frequency, this method is far more sensitive to small changes of parameters, and yields very precise estimates on them.
The exact parameters we fitted were the difference from F0 in units of $10^{-X}$\,s, depending on the observation length, the difference from F1 in units of $10^{-Y}$\,Hz/s (where $X,Y$ were chosen as values close to the order of magnitude of the known errors on the parameters), and the difference from TASC in seconds.
This was done to avoid incurring in numerical errors due to the small steps involved in some parameters.
We set flat priors for all parameters: $0.5 < f < 1$ Hz, $|\dot{f}| < 10^{-7}$ and $\Delta\tasc < \Porb$.
We first used the \texttt{scipy.optimize.minimize} function to minimize the negative log-likelihood and determine an approximate starting solution.
Then, values around this solution was used to initialize a Markov Chain Monte Carlo (MCMC) sampler as implemented in the \texttt{emcee} \citep{2013PASP..125..306F} library.
Since the analysis took a significant time (up to 2 s per iteration in the larger datasets), we followed the instructions in the \texttt{emcee} documentation to interrupt the sampling once the iterations had reached 200 times the the ``autocorrelation time'' $\tau$, more than the recommended 50 for additional robustness.
$\tau$ itself was calculated every 100 steps of the chain.
The number of steps used in the chains varied between 3,000 and 100,000 depending mostly on the length of the observation and the number of photons, with longer observations requiring fewer steps (because of the reduced correlations between parameters).
We used the 3-30\,keV or 8-30\,keV  energy range depending on which range yielded the highest power in the Rayleigh search.
This allowed to estimate the posterior distribution on the parameters, and their uncertainties.
The posterior distributions are generally well-behaved, with reasonably (sometimes slightly skewed due to the correlation between the parameters) bell-shaped distributions.
We determined 1-$\sigma$ error bars on the parameters by looking at the 16\% and 84\% percentiles.
The results are summarized in \tabref{tab:obs}, and the detection in ObsID 30101045002 is shown in \figref{fig:search} as an example.

\subsection{Orbital decay}\label{sec:decay}\label{sec:orbit}
To describe a change of the orbital period over time, it is customary to measure the time that a star reaches a particular phase of the orbit, and compare it with the expected time given the previous orbital solution.
For circular orbits with no eclipses, it is common to use one of the two intersections between the orbit and the plane perpendicular to the line of sight passing through the center of mass of the binary system.
These points of the orbit are called nodes;
the \textit{ascending node} is the node that the pulsar crosses when moving away from the observer.
The expected time of passage at the ascending node after $n$ orbits $T_{\rm asc,n}$ (for simplicity, we drop the \textit{asc} when $n$ or other indices are present), in the presence of an orbital derivative, can be expressed as (cfr. \citealt{kelleyDeterminationOrbitGX1980,falangaEphemerisOrbitalDecay2015}):
\begin{equation}\label{eq:Tn}
    T_n = T_0 + n\,\Porb + \frac{1}{2}\,n^2\,\Porb\,\Porbdot +\,\frac{1}{6}\,n^3\Porb^2\,\ddot{P}_{\rm orb} +\dots
\end{equation}
By using a previously determined ephemeris as a baseline, we can measure the delay of the measured $\tasc$ from the expected one.
When plotting this delay, offsets indicate a shift in $\tasc$, linear trends an uncertainty of $\Porb$, and parabolic trends an orbital period derivative:
\begin{equation}\label{eq:deltaTn}
    \delta T_n(t) = \delta \tasc + \frac{t - \tasc}{\Porb}\delta{\Porb} + \frac{1}{2}\frac{\Porbdot}{\Porb}\left(t - \tasc\right)^2,
\end{equation}
where we substituted $n=(t-\tasc)/\Porb$.

\begin{figure*}[ht]
    \centering
    \includegraphics[width=0.48\linewidth]{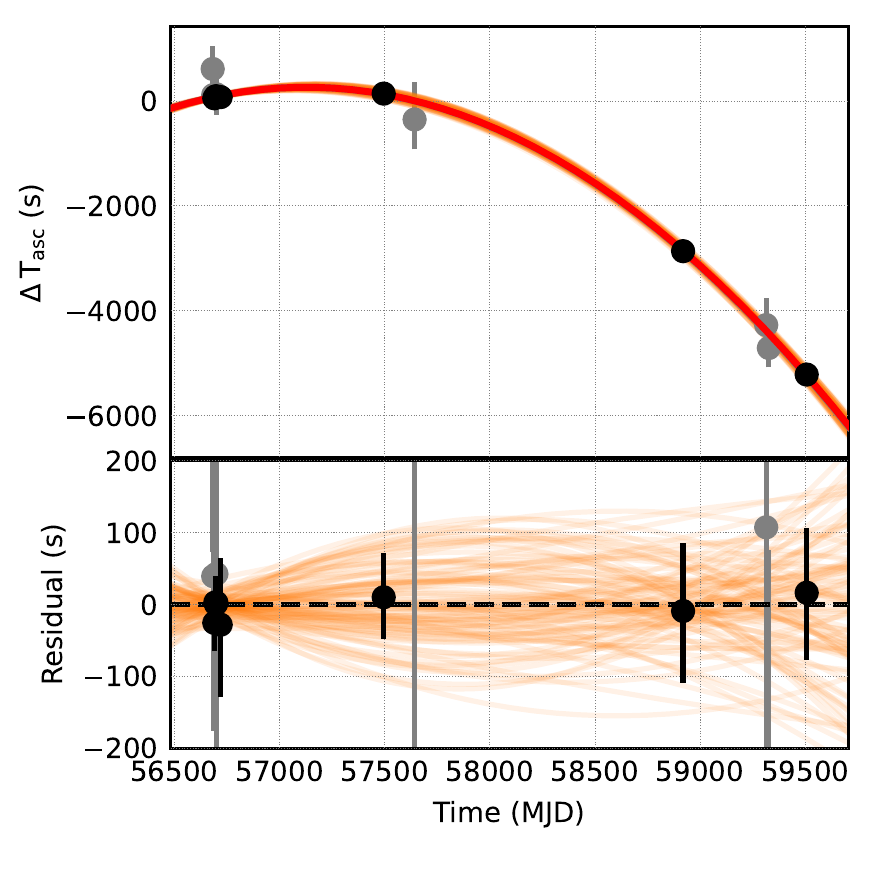}
    \includegraphics[width=0.48\linewidth]{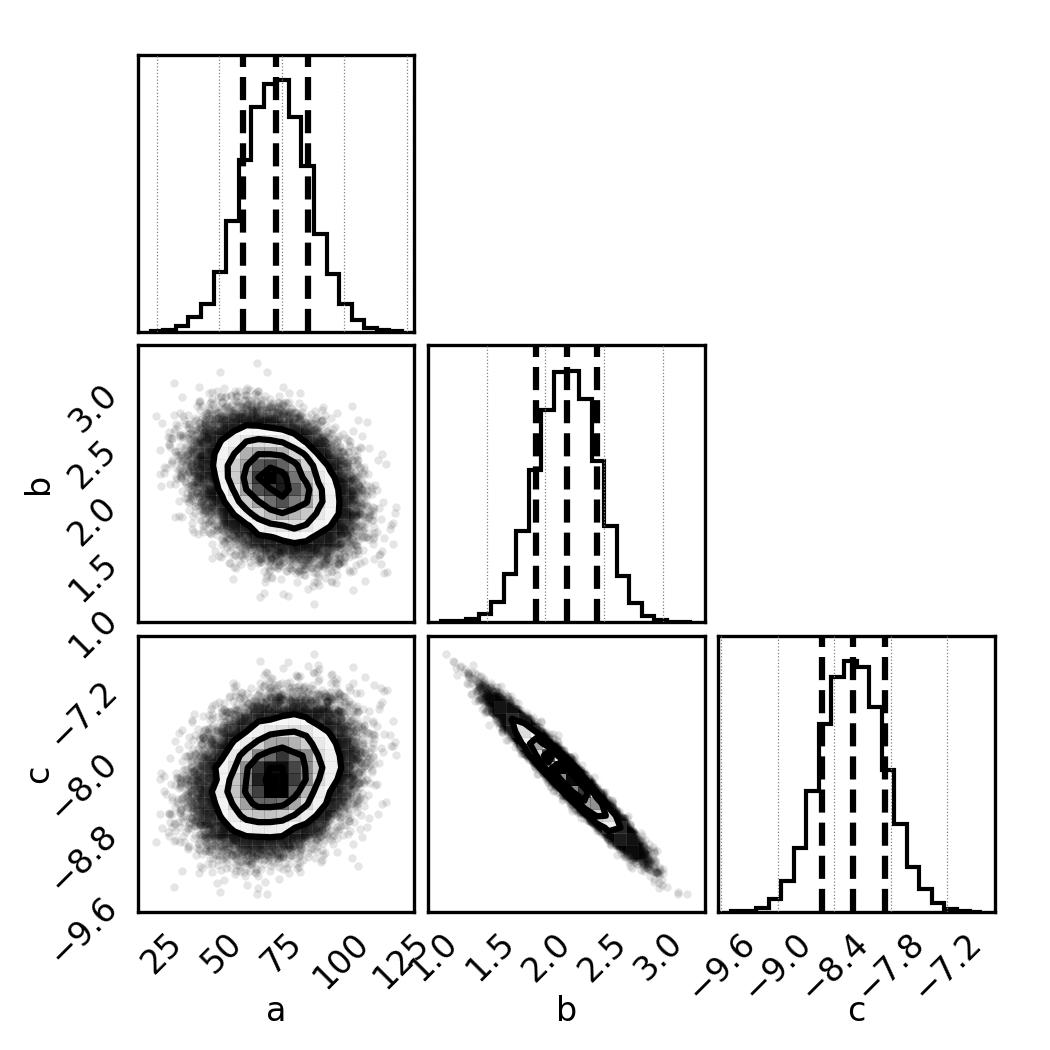}
    \caption{(Left) Orbital decay in \Mtwo, measured through the delay of the $\tasc$ parameter (time of passage through the ascending node) from Equation~\ref{eq:deltaTn}. Semi-transparent orange lines indicate possible $\sim$1000 quadratic solutions, coming from MCMC sampling (described in \secref{sec:orbit}).
    Grey points are lower-quality measurements (uncertainty in \tasc larger than 200 s due to high orbital-spin parameter correlation).
    (Right) Corner plot of the posterior distribution of the parameters  of the orbital decay with Equation~\ref{eq:map}, sampled with MCMC. Vertical dashed lines show the 16\%, 50\%, and 84\% percentiles.
    }
    \label{fig:decay}\label{fig:mcmc}
\end{figure*}

Using the new $T_{\rm asc}$ values, we infer the orbital decay of \Mtwo\ using a Bayesian model.

The following equation serves as the orbital evolution model:
\begin{equation}\label{eq:map}
    \Delta \tasc = a +  \frac{b}{\Porb} (t - \tasc) + 0.5\cdot\msix\,c\, \frac{86400}{365.25} (t - \tasc)^2
\end{equation}

where $t$, $\tasc$ and $\Porb$ are expressed in days, $\Delta \tasc$ is the delay of $\tasc$ in seconds, $a$ is a correction to $\tasc$ in seconds, $b$ is a correction to $\Porb$ in seconds and $c$ is the new value of $\Porbdot/\Porb$ in units of $\msixyrmone$.
The baseline solution from B20 was $\tasc_{,\rm B20}=\mathrm{MJD}\,56682.0661$, $P_{\rm orb,B20}=2.532948$\,d, and $\Porbdot=0$\,s~s$^{-1}$.
Priors for $a$, $b$ and $c$ were uniform between $\pm10^6$; in checks, we found that the width of the prior has no significant effect on our posterior inference.

\begin{table}[ht]
    \centering
        \begin{tabular}{ccc}
         \hline
        Parameter & Unit & Value (uncert)  \\
        \hline
        $\Porb$ & d & 2.5329733(30) \\
        $\Porbdot$ & s~s$^{-1}$ & $-5.69(24) \cdot 10^{-8}$ \\
        $\Porbdot/\Porb$ & yr$^{-1}$ & $-8.20(34)\cdot \msix$ \\
        $a \sin i$ & l-sec & 22.218(5) \\
        $\tasc$ & MJD & 56682.06694(15)\\
        $e$ & & $<0.0015$ (3-$\sigma$ u.l.)\\
        \hline
        \end{tabular}
    \caption{Updated orbital parameters for \Mtwo, as determined in this work.}
    \label{tab:eph}
\end{table}

We first performed a Maximum-A-Posteriori fit with a standard Gaussian likelihood, allowing for asymmetric error bars. The solution served as an initialization of a Markov Chain Monte Carlo (MCMC) sampler using \texttt{emcee} as before. Using 32 walkers, we ran the chains for 20000 steps.
We calculated the autocorrelation ``time'', which was at most 46 steps. We thinned the chain by a factor 23 (half the autocorrelation length) and discarded 920 steps (20 times the autocorrelation length) as burn-in.
The resulting marginal posterior probability distributions are plotted using the \texttt{corner} library \citep{corner} in Figure ~\ref{fig:mcmc}.

We find posterior means and credible intervals of $a=72(13)$, $b=2.18(26)$, and $c=-8.20(34)$. Using these values, we corrected the orbital parameters as $\tasc=T_{0,\rm B20} + a\,\mathrm{sec}$, $\Porb=P_{\rm orb,B20} + b\,\mathrm{sec}$, $\Porbdot/\Porb=c\cdot\msix {\rm yr}^{-1}$ to obtain the values in Table~\ref{tab:eph}.

Finally, we fixed the orbital parameters and we re-ran a final accelerated search for pulsations in all ObsIDs using the Rayleigh test, yielding the results in \tabref{tab:obs}.

\section{Discussion}

Over the years, many models have been proposed to describe the interaction between the plasma in the disk and the magnetic field lines of an accreting pulsar \citep{ghoshDiskAccretionMagnetic1978,wangLocationInnerRadius1996,chashkina2017}.
Despite large differences in the treatment of the details of this interaction, these models make estimates for the magnetic field within $\sim$one order of magnitude when the inner radius and the mass accretion rate are fixed \citep{xuMagneticFieldUltraluminous2017,chenStudyingMagneticFields2021,erkutMagneticFieldsBeaming2020}, if one can assume \textit{spin equilibrium}: a regime where the outward pressure from the rotating magnetic field balances almost exactly the ram pressure from the infalling matter.

Until now, different groups have used the observed luminosity as a proxy for the mass accretion rate, and this produced very different estimates depending on the assumption of the beaming fraction.
In addition, different works used different assumptions on the position of the inner radius, with the high-magnetic field models assuming spin equilibrium \citep{eksiUltraluminousXraySource2015,tsygankovPropellerEffectAction2016,dallossoNuSTARJ09555169402015} and the beaming models being incompatible with it \citep{kingPulsingULXsTip2017}.

In this work, we produce robust evidence in favor of spin equilibrium, and we measure an orbital decay that might provide an independent estimate of mass transfer, as we are going to discuss below.

\subsection{Spin equilibrium}\label{sec:equilibrium}

Thanks to the new detections listed above, we found that for at least part of the time between 2016 and 2020 the pulsar continued to spin down (slow down its rotation) as reported by B20, because the frequency ($\sim0.721$\,Hz) observed in 2020 was lower than observed in 2016 ($\sim0.723$\,Hz).
However, since then, the neutron star appears to be alternating phases of spin up and spin down around $\sim0.721$\,Hz.
In at least one observation in 2016 and probably in another in 2021, the pulsar was spinning down while accreting (see Table~\ref{tab:obs}).
In summary, the spin evolution of \Mtwo\ strongly points to a situation of spin equilibrium.
In this condition, spin up and spin down can be produced with small changes of accretion rate \citep{dangeloAccretionDiscsTrapped2012}, and it is possible to confidently estimate the magnetic field of the neutron star, by equating the analytical formulas for the inner radius $\rin$ to the corotation radius $\rco$, at which the angular velocity of the matter in the disk equals the one of the star (see \secref{sec:radii}).

Being close to spin equilibrium also implies that a relatively small drop of mass accretion rate could trigger the so-called ``propeller'' regime \citep{illarionovWhyNumberGalactic1975}, where the rotating, highly-magnetized pulsar is able to swipe away the infalling matter.
During the transition to this regime, it is possible to still have accretion, albeit discontinuous \citep{romanovaPropellerRegimeDisk2004}, up to a point where accretion is stopped altogether leaving only the disk and the ejected matter as sources of X-ray flux.
Based on a possible bimodal distribution of the fluxes of \Mtwo, \citet{tsygankovPropellerEffectAction2016} claimed that the observed low states in \Mtwo were evidence of this transition.
It is not clear, at the moment, if this is compatible with the observed $\sim60$-day periodicity of the low states \citep{brightman60DaySuperorbital2019}, which would imply a \textit{periodic} decrease of mass transfer, difficult to reconcile with the very low eccentricity of the system.

\subsection{Is it mass transfer?}
\begin{figure}[ht]
    \centering
    \includegraphics[width=\linewidth]{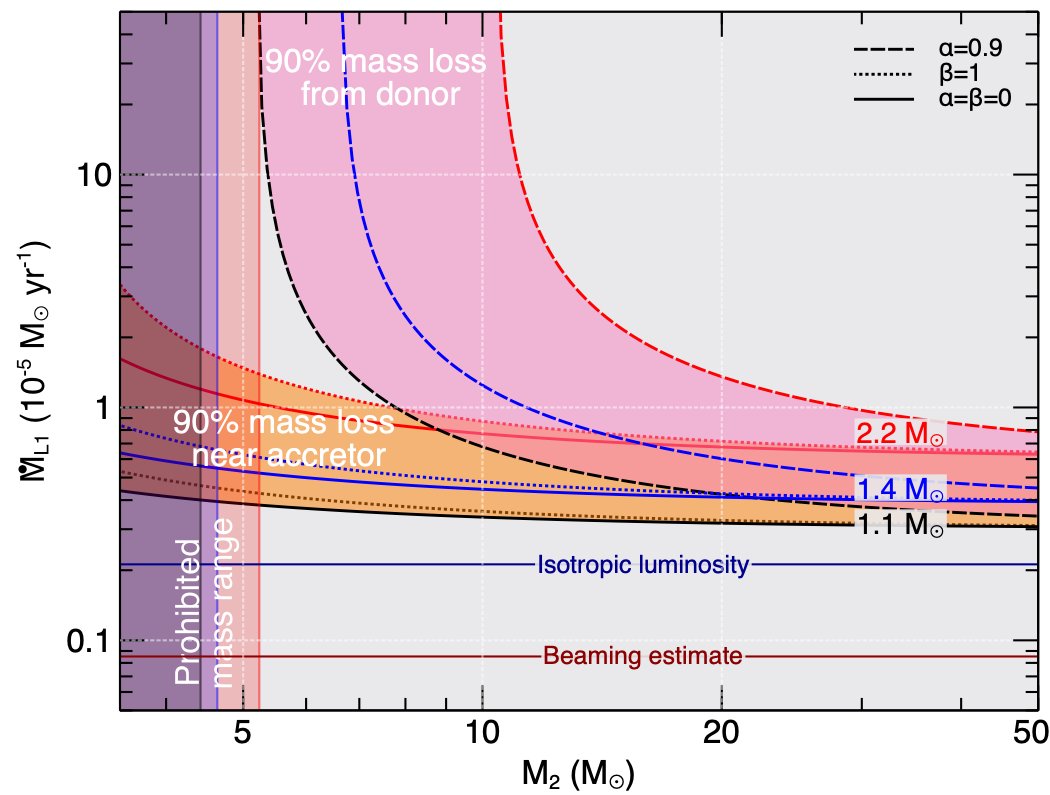}
    \caption{Mass transfer rate towards the accretor (through L1), versus donor mass, given the measured orbital period decay.
    We estimate the mass transfer rate for different mechanisms of mass loss and different accretor masses.
    Solid lines indicate conservative mass transfer, dashed lines indicate 90\% isotropic mass loss from the donor, and dotted lines indicate 100\% mass loss from the proximity of the accretor.
    The latter is only a small correction to the conservative case, while mass loss from the donor implies a much larger mass transfer rate in order to produce the observed orbital decay.
    Vertical bands on the left show the limit donor masses given the absence of eclipses, for different accretor masses (same color coding).}
    \label{fig:mdot}
\end{figure}

The observed orbital decay is compatible with the mass transfer from a more massive donor star to a neutron star (\citealt{taurisFormationEvolutionCompact2006}, see \secref{sec:transfer}).
Assuming a pulsar mass $M_p=1.4\msun$ and a donor mass $M_d=8\msun$ (which corresponds to the mean of the probability distributions of masses, see \secref{sec:donor}), it is straightforward to estimate the mass transfer rate from the observed orbital decay, assuming conservative mass transfer, as $\Mdot_d \approx -4.7\cdot\msix\msunyr$.
This corresponds to $\sim200$ times the Eddington limit, assuming an Eddington mass accretion rate corresponding to $\Mdot_{\rm Edd}\approx1.5\cdot10^{18}\,{\rm g~s^{-1}}\approx2.4\cdot10^{-8}\,{\msun \rm yr^{-1}}$.
This is the mass that the donor transfers into the Roche lobe of the neutron star.
This mass exchange exceeds both that inferred from the apparent bolometric luminosity of the source (which is at most $\sim100$ times the Eddington limit, see B14), or the one inferred from beaming scenarios (36 times Eddington, \citealt{kingPulsingULXsTip2017}).
See \figref{fig:mdot} for details.
It is possible that part of this matter leaves the system through fast winds launched from the super-Eddington disk \citep{pintoUltrafastOutflowsUltraluminous2016,kosecEvidenceVariableUltrafast2018}.
This mass loss happens from the vicinity of the accretor, and its specific angular momentum is such that the effect on the orbit is similar to the conservative case (see Appendix~\ref{sec:transfer}).
On the other hand, it decreases the amount of matter that accretes onto the neutron star, which is a viable explanation for the slightly lower luminosity observed.
Isotropic mass loss from the donor, instead, as one would expect from stellar winds, would have the opposite effect, expanding the orbit.
Additional mass loss from the outer disk in the form of slow winds \citep[e.g.][]{middletonThermallyDrivenWinds2022} would represent an intermediate case, carrying away specific angular momentum somewhere between that at the position of the neutron star and that at L1.
Therefore, the estimate above is a \textit{lower limit} to the mass transfer rate.
Another possibility, involving mass loss from L2 forming a circumbinary disk, is discussed below.

If the observed orbital decay really is due to mass transfer, we can fix these two important variables, leading to an interpretation of \Mtwo\ being a highly magnetized neutron star ($B>10^{13}$\,G) with any of these models (see also Appendix~\ref{sec:magnetic}), and exclude beaming as a primary amplifier of ULX emission (also see \citealt{vasilopoulosChandraProbesXRay2021} for further evidence in this sense).

\subsection{Alternative models}
\subsubsection{Synchronization and circularization}
The value of orbital derivative found in this work is larger, albeit only by a few times, than those observed in much less luminous X-ray pulsars such as SMC X-1, LMC X-4, and Cen X-3 ($-3.5$, $-1.0$, and $-1.8\cdot\msixyrmone$ respectively, see \citealt{falangaEphemerisOrbitalDecay2015}).
The slight mismatch that we find in \Mtwo between the mass transfer inferred from the orbit and that inferred from the luminosity becomes a factor $\sim$20 for SMC X-1.
This has led Falanga and many other authors \citep[e.g.]{levineOrbitalDecayLMC2000,levineDiscoveryOrbitalDecay1993,chernovOrbitalDecayHighMass2020} to disregard mass transfer as the primary engine for orbital decay in these systems.
At this point, we cannot exclude that whatever process believed to be in place in those systems (like viscous processes producing the circularization of an elliptical orbit or the synchronization of the star's rotation with the orbit, e.g.
\citealt{falangaEphemerisOrbitalDecay2015} and references therein; \citealt{chernovOrbitalDecayHighMass2020}) is at work in \Mtwo.
However, we do stress that these HMXBs are likely accreting in very different regimes, possibly from focused winds and not from a Roche-Lobe overflow, as would instead be expected from a super-Eddington source.

\subsubsection{Circumbinary disk}
The observed orbital decay is, in principle, compatible with an equatorial circumbinary disk launched by the second Lagrangian point L2 (\citealt{taurisFormationEvolutionCompact2006}, see also Appendix~\ref{sec:transfer}.
However, this only happens in situations where the donor star inflates well beyond its Roche Lobe, or via a fast and stable wind, and its onset quickly leads to unstable orbital decay and common envelope \citep{misraOriginPulsatingUltraluminous2020}.
\citet{lu2022} study in detail the conditions for this phenomenon, finding that it should be important above $\mdot\sim10^{-4}\msunyr$.
SS433, a possible ULX analog in our Galaxy \citep{fabrikaPropertiesSS433Ultraluminous2006, middletonNuSTARRevealsHidden2021}, might be undergoing such a process.
However, in that case, the accretor is believed to be a stellar-mass black hole \citep{blundellSS433Observation2008} and the mass ratio is $\sim 1$, and this process leads to the \textit{expansion} of the orbit \citep{Cherepashchuk+2021}, stabilizing the mass transfer.

\section{Conclusion}
The detection of orbital decay in \Mtwo is a key milestone to understand the evolution of this system and, possibly, of all low-orbital period PULXs like NGC 5907 ULX1 \citep{israelAccretingPulsarExtreme2017} and M51 ULX-7 \citep{rodriguezcastilloDiscoveryPulsarDay2020}.

We argue that the decay is driven by mass transfer: the implied mass transfer is only a factor $\sim2$ above the one inferred from the luminosity, and this can easily be explained by a slightly lower efficiency or a massive outflow such as those observed in other ULXs (but undetectable in \Mtwo due to source confusion in the M82 field).
Currently, we cannot exclude that phenomena such as the synchronization of the donor rotation with the orbit and/or the circularization of the orbit are contributing to the observed decay, which is only a few times higher than that of the eclipsing HMXBs from  \citet{falangaEphemerisOrbitalDecay2015}.
Note however that this source has a quite tight upper limit on the eccentricity (see Appendix~\ref{sec:ecc}) and the accretion is very likely through Roche Lobe overflow, at a much higher rate than the sample from \citet{falangaEphemerisOrbitalDecay2015}, making the timescale for synchronization faster.
If these phenomena are in place, it is likely that we are witnessing a very short-lived phase of the evolution of this binary system.

Regardless of the exact driver of the observed orbital decay, our measurement informs the theoretical study of ULX progenitors.
At the moment, the evolutionary scenarios able to produce a ULX seem to often lead to common envelope, with a relatively short phase of extreme mass transfer, in particular for donor masses in the lower mass of the allowed ranges for \Mtwo \citep{taurisFormationEvolutionCompact2006,misraOriginPulsatingUltraluminous2020}.
Stabilizing mechanisms, such as mass loss from the donor, are often invoked to increase the lifespan of ULXs, provided that the envelope is radiative.

We encourage further theoretical studies on the evolution of binary systems, to understand the conditions in which an orbital decay such as the one we observe can be produced.
Future missions with instruments at higher throughput, like Athena \citep{barconsAthenaESAXray2017}, will help detect pulsations and perform similar studies in many more ULXs.
For \Mtwo and extragalactic pulsars in general, which have hard pulsations and are often found in crowded fields, hard imagers with high angular resolution and good timing capabilities, like the proposed NASA probe HEX-P \citep{madsenHEXPHighEnergyXray2019}, would be excellent.
Timing-devoted missions with large collecting area, such as the Chinese-Italian eXTP \citep{zhangEXTPEnhancedXray2016} or the proposed NASA probe STROBE-X \citep{raySTROBEXProbeclassMission2018}, will allow sensitive searches for pulsations and timing studies in ULXs, provided that they are sufficiently isolated.

\begin{acknowledgments}

The authors wish to thank Victoria Grinberg, W{\l}odek Klu\'zniak and Alessandro Ridolfi for useful discussions, and the staff at the NuSTAR Science Operations Center at Caltech for the help in scheduling the observations and the frequent clock correction file updates, that allowed a prompt analysis of the data.
We would also wish to thank the anonymous referee, and the three referees of a previous submission, who provided very insightful feedback that allowed to substantially improve the quality of the analysis.
MB was funded in part by PRIN TEC
INAF 2019 ``SpecTemPolar! -- Timing analysis in the era of high-throughput photon detectors''.
MH is supported by an ESO fellowship.
GLI and MB  acknowledge  funding  from  the  Italian MIUR PRIN grant 2017LJ39LM.
ADJ was funded in part by the Chandra grant 803-0000-716015-404H00-6100-2723-4210-40716015HH83121.
JP was supported by the grant 14.W03.31.0021 of the Ministry of Science and Higher Education of the Russian Federation and the Academy of Finland grant 333112.
DJW acknowledges support from STFC in the form of an Ernest Rutherford Fellowship.
HPE acknowledges support under NASA Contract No. NNG08FD60C.

All the analysis of this Paper was done using open-source software: Astropy, Stingray, HENDRICS, PINT, emcee, corner, scinum, and can easily be verified  using the solutions in Table~\ref{tab:eph} and~\ref{tab:obs}.
The implementation of the \citet{pletschGammaRayTimingRedback2015} method can be found in the github repository \url{https://github.com/matteobachetti/ell1fit}.
Figures were produced using the Matplotlib library and the Veusz software.
The data used for this work come from the \nustar and \xmm missions and are usually held private for one year, and made public on the High Energy Astrophysics Science archive (HEASARC) and the XMM-Newton Science Archive (XSA) afterwise. \nustar is a Small Explorer mission led by Caltech and managed by JPL for NASA's Science Mission Directorate in Washington. \nustar was developed in partnership with the Danish Technical University and the Italian Space Agency (ASI). The spacecraft was built by Orbital Sciences Corp., Dulles, Virginia. 
\xmm is an ESA science mission with instruments and contributions directly funded by ESA Member States and NASA

\end{acknowledgments}

%

\vspace{5mm}
\facilities{NuSTAR}


\software{astropy \citep{astropycollaborationAstropyProjectBuilding2018a},
          Stingray \citep{huppenkothenStingraySpectraltimingSoftware2016},
          HENDRICS \citep{bachettiHENDRICSHighENergy2018},
          PINT \citep{luoPINTModernSoftware2021},
          scipy \citep{2020SciPy-NMeth},
          numpy \citep{harris2020array},
          numba \citep{numba2015},
          Veusz\footnote{https://veusz.github.io/},
          Matplotlib \citep{Hunter:2007},
          corner \citep{corner},
          emcee \citep{2013PASP..125..306F}
          }

\clearpage


\bibliography{m82_orbit}{}

\begin{thebibliography}{}
\expandafter\ifx\csname natexlab\endcsname\relax\def\natexlab#1{#1}\fi
\providecommand{\url}[1]{\href{#1}{#1}}
\providecommand{\dodoi}[1]{doi:~\href{http://doi.org/#1}{\nolinkurl{#1}}}
\providecommand{\doeprint}[1]{\href{http://ascl.net/#1}{\nolinkurl{http://ascl.net/#1}}}
\providecommand{\doarXiv}[1]{\href{https://arxiv.org/abs/#1}{\nolinkurl{https://arxiv.org/abs/#1}}}

\bibitem[{Abarca {et~al.}(2021)Abarca, Parfrey, \&
  Klu{\'z}niak}]{abarcaBeamedEmissionNeutronstar2021}
Abarca, D., Parfrey, K., \& Klu{\'z}niak, W. 2021, The Astrophysical Journal,
  917, L31, \dodoi{10.3847/2041-8213/ac1859}

\bibitem[{{Astropy Collaboration} {et~al.}(2018){Astropy Collaboration},
  {Price-Whelan}, Sip{\H o}cz, G{\"u}nther, Lim, Crawford, Conseil, Shupe,
  Craig, Dencheva, Ginsburg, VanderPlas, Bradley, {P{\'e}rez-Su{\'a}rez}, {de
  Val-Borro}, Aldcroft, Cruz, Robitaille, Tollerud, Ardelean, Babej, Bach,
  Bachetti, Bakanov, Bamford, Barentsen, Barmby, Baumbach, Berry, Biscani,
  Boquien, Bostroem, Bouma, Brammer, Bray, Breytenbach, Buddelmeijer, Burke,
  Calderone, Cano~Rodr{\'i}guez, Cara, Cardoso, Cheedella, Copin, Corrales,
  Crichton, D'Avella, Deil, Depagne, Dietrich, Donath, Droettboom, Earl, Erben,
  Fabbro, Ferreira, Finethy, Fox, Garrison, Gibbons, Goldstein, Gommers, Greco,
  Greenfield, Groener, Grollier, Hagen, Hirst, Homeier, Horton, Hosseinzadeh,
  Hu, Hunkeler, Ivezi{\'c}, Jain, Jenness, Kanarek, Kendrew, Kern, Kerzendorf,
  Khvalko, King, Kirkby, Kulkarni, Kumar, Lee, Lenz, Littlefair, Ma, Macleod,
  Mastropietro, McCully, Montagnac, Morris, Mueller, Mumford, Muna, Murphy,
  Nelson, Nguyen, Ninan, N{\"o}the, Ogaz, Oh, Parejko, Parley, Pascual, Patil,
  Patil, Plunkett, Prochaska, Rastogi, Reddy~Janga, Sabater, Sakurikar,
  Seifert, Sherbert, {Sherwood-Taylor}, Shih, Sick, Silbiger, Singanamalla,
  Singer, Sladen, Sooley, Sornarajah, Streicher, Teuben, Thomas, Tremblay,
  Turner, Terr{\'o}n, {van Kerkwijk}, {de la Vega}, Watkins, Weaver, Whitmore,
  Woillez, Zabalza, \& {Astropy
  Contributors}}]{astropycollaborationAstropyProjectBuilding2018a}
{Astropy Collaboration}, {Price-Whelan}, A.~M., Sip{\H o}cz, B.~M., {et~al.}
  2018, The Astronomical Journal, 156, 123, \dodoi{10.3847/1538-3881/aabc4f}

\bibitem[{Bachetti(2018)}]{bachettiHENDRICSHighENergy2018}
Bachetti, M. 2018, Astrophysics Source Code Library, ascl:1805.019

\bibitem[{Bachetti {et~al.}(2021)Bachetti, Pilia, Huppenkothen, Ransom,
  Curatti, \& Ridolfi}]{bachettiExtendingStatisticsGeneric2021}
Bachetti, M., Pilia, M., Huppenkothen, D., {et~al.} 2021, ApJ, 909, 33,
  \dodoi{10.3847/1538-4357/abda4a}

\bibitem[{Bachetti {et~al.}(2013)Bachetti, Rana, Walton, Barret, Harrison,
  Boggs, Christensen, Craig, Fabian, F{\"u}rst, Grefenstette, Hailey,
  Hornschemeier, Madsen, Miller, Ptak, Stern, Webb, \&
  Zhang}]{bachettiUltraluminousXRaySources2013}
Bachetti, M., Rana, V., Walton, D.~J., {et~al.} 2013, ApJ, 778, 163,
  \dodoi{10.1088/0004-637X/778/2/163}

\bibitem[{Bachetti {et~al.}(2014)Bachetti, Harrison, Walton, Grefenstette,
  Chakrabarty, F{\"u}rst, Barret, Beloborodov, Boggs, Christensen, Craig,
  Fabian, Hailey, Hornschemeier, Kaspi, Kulkarni, Maccarone, Miller, Rana,
  Stern, Tendulkar, Tomsick, Webb, \&
  Zhang}]{bachettiUltraluminousXraySource2014}
Bachetti, M., Harrison, F.~A., Walton, D.~J., {et~al.} 2014, Nat., 514, 202,
  \dodoi{10.1038/nature13791}

\bibitem[{Bachetti {et~al.}(2020)Bachetti, Maccarone, Brightman, Brumback,
  F{\"u}rst, Harrison, Heida, Israel, Middleton, Tomsick, Webb, \&
  Walton}]{bachettiAllOnceTransient2020}
Bachetti, M., Maccarone, T.~J., Brightman, M., {et~al.} 2020, ApJ, 891, 44,
  \dodoi{10.3847/1538-4357/ab6d00}

\bibitem[{Barcons {et~al.}(2017)Barcons, Barret, Decourchelle, {den Herder},
  Fabian, Matsumoto, Lumb, Nandra, Piro, Smith, \&
  Willingale}]{barconsAthenaESAXray2017}
Barcons, X., Barret, D., Decourchelle, A., {et~al.} 2017, Astronomische
  Nachrichten, 338, 153, \dodoi{10.1002/asna.201713323}

\bibitem[{Blundell {et~al.}(2008)Blundell, Bowler, \&
  Schmidtobreick}]{blundellSS433Observation2008}
Blundell, K.~M., Bowler, M.~G., \& Schmidtobreick, L. 2008, The Astrophysical
  Journal, 678, L47, \dodoi{10.1086/588027}

\bibitem[{Brice {et~al.}(2021)Brice, Zane, Turolla, \&
  Wu}]{briceSupereddingtonEmissionAccreting2021}
Brice, N., Zane, S., Turolla, R., \& Wu, K. 2021, Monthly Notices of the Royal
  Astronomical Society, 504, 701, \dodoi{10.1093/mnras/stab915}

\bibitem[{Brightman {et~al.}(2019)Brightman, Harrison, Bachetti, Xu, F{\"u}rst,
  Walton, Ptak, Yukita, \& Zezas}]{brightman60DaySuperorbital2019}
Brightman, M., Harrison, F.~A., Bachetti, M., {et~al.} 2019, The Astrophysical
  Journal, 873, 115, \dodoi{10.3847/1538-4357/ab0215}

\bibitem[{Buccheri {et~al.}(1983)Buccheri, Bennett, Bignami, Bloemen,
  Boriakoff, Caraveo, Hermsen, Kanbach, Manchester, Masnou,
  {Mayer-Hasselwander}, Ozel, Paul, Sacco, Scarsi, \&
  Strong}]{buccheriSearchPulsedGammaray1983}
Buccheri, R., Bennett, K., Bignami, G.~F., {et~al.} 1983, A\&A, 128, 245

\bibitem[{Chashkina {et~al.}(2017)Chashkina, Abolmasov, \&
  Poutanen}]{chashkina2017}
Chashkina, A., Abolmasov, P., \& Poutanen, J. 2017, Monthly Notices of the
  Royal Astronomical Society, 470, 2799, \dodoi{10.1093/mnras/stx1372}

\bibitem[{Chashkina {et~al.}(2019)Chashkina, Lipunova, Abolmasov, \&
  Poutanen}]{chashkinaSuperEddingtonAccretionDiscs2019}
Chashkina, A., Lipunova, G., Abolmasov, P., \& Poutanen, J. 2019, Astronomy and
  Astrophysics, 626, A18, \dodoi{10.1051/0004-6361/201834414}

\bibitem[{Chen {et~al.}(2021)Chen, Wang, \&
  Tong}]{chenStudyingMagneticFields2021}
Chen, X., Wang, W., \& Tong, H. 2021, Journal of High Energy Astrophysics, 31,
  1, \dodoi{10.1016/j.jheap.2021.04.002}

\bibitem[{Cherepashchuk {et~al.}(2021)Cherepashchuk, Belinski, Dodin, \&
  Postnov}]{Cherepashchuk+2021}
Cherepashchuk, A.~M., Belinski, A.~A., Dodin, A.~V., \& Postnov, K.~A. 2021,
  Monthly Notices of the Royal Astronomical Society, 507, L19,
  \dodoi{10.1093/mnrasl/slab083}

\bibitem[{Chernov(2020)}]{chernovOrbitalDecayHighMass2020}
Chernov, S.~V. 2020, Astron. Rep., 64, 425, \dodoi{10.1134/S1063772920050017}

\bibitem[{Dall'Osso {et~al.}(2015)Dall'Osso, Perna, \&
  Stella}]{dallossoNuSTARJ09555169402015}
Dall'Osso, S., Perna, R., \& Stella, L. 2015, MNRAS, 449, 2144,
  \dodoi{10.1093/mnras/stv170}

\bibitem[{D'Angelo \& Spruit(2012)}]{dangeloAccretionDiscsTrapped2012}
D'Angelo, C.~R., \& Spruit, H.~C. 2012, Monthly Notices of the Royal
  Astronomical Society, 420, 416, \dodoi{10.1111/j.1365-2966.2011.20046.x}

\bibitem[{Eggleton(1983)}]{eggletonApproximationsRadiiRoche1983}
Eggleton, P.~P. 1983, ApJ, 268, 368, \dodoi{10.1086/160960}

\bibitem[{Ek{\c s}i {et~al.}(2015)Ek{\c s}i, Anda{\c c}, {\c C}{\i}k{\i}nto{\u
  g}lu, Gen{\c c}ali, G{\"u}ng{\"o}r, \&
  {\"O}ztekin}]{eksiUltraluminousXraySource2015}
Ek{\c s}i, K.~Y., Anda{\c c}, {\textbackslash}.~C., {\c C}{\i}k{\i}nto{\u g}lu,
  S., {et~al.} 2015, MNRAS Let., 448, L40, \dodoi{10.1093/mnrasl/slu199}

\bibitem[{Erkut {et~al.}(2020)Erkut, T{\"u}rko{\u g}lu, Ek{\c s}i, \&
  Alpar}]{erkutMagneticFieldsBeaming2020}
Erkut, M.~H., T{\"u}rko{\u g}lu, M.~M., Ek{\c s}i, K.~Y., \& Alpar, M.~A. 2020,
  The Astrophysical Journal, 899, 97, \dodoi{10.3847/1538-4357/aba61b}

\bibitem[{Fabrika {et~al.}(2006)Fabrika, Karpov, Abolmasov, Sholukhova, \&
  Fabbiano}]{fabrikaPropertiesSS433Ultraluminous2006}
Fabrika, S., Karpov, S., Abolmasov, P., Sholukhova, O., \& Fabbiano, G. 2006,
  IAU, 230, 278, \dodoi{10.1017/S1743921306008441}

\bibitem[{Fabrika {et~al.}(2021)Fabrika, Atapin, Vinokurov, \&
  Sholukhova}]{fabrikaUltraluminousXRaySources2021}
Fabrika, S.~N., Atapin, K.~E., Vinokurov, A.~S., \& Sholukhova, O.~N. 2021,
  Astrophys. Bull., 76, 6, \dodoi{10.1134/S1990341321010077}

\bibitem[{Falanga {et~al.}(2015)Falanga, Bozzo, Lutovinov, {Bonnet-Bidaud},
  Fetisova, \& Puls}]{falangaEphemerisOrbitalDecay2015}
Falanga, M., Bozzo, E., Lutovinov, A., {et~al.} 2015, Astronomy and
  Astrophysics, 577, A130, \dodoi{10.1051/0004-6361/201425191}

\bibitem[{Foreman-Mackey(2016)}]{corner}
Foreman-Mackey, D. 2016, The Journal of Open Source Software, 1, 24,
  \dodoi{10.21105/joss.00024}

\bibitem[{{Foreman-Mackey} {et~al.}(2013){Foreman-Mackey}, {Hogg}, {Lang}, \&
  {Goodman}}]{2013PASP..125..306F}
{Foreman-Mackey}, D., {Hogg}, D.~W., {Lang}, D., \& {Goodman}, J. 2013, PASP,
  125, 306, \dodoi{10.1086/670067}

\bibitem[{Frank {et~al.}(2002)Frank, King, \&
  Raine}]{frankAccretionPowerAstrophysics2002}
Frank, J., King, A., \& Raine, D.~J. 2002, Accretion {{Power}} in
  {{Astrophysics}}: {{Third Edition}} ({Accretion Power in Astrophysics})

\bibitem[{Ghosh \& Lamb(1978)}]{ghoshDiskAccretionMagnetic1978}
Ghosh, P., \& Lamb, F.~K. 1978, ApJ, 223, L83, \dodoi{10.1086/182734}

\bibitem[{Gladstone {et~al.}(2009)Gladstone, Roberts, \&
  Done}]{gladstoneUltraluminousState2009}
Gladstone, J.~C., Roberts, T.~P., \& Done, C. 2009, MNRAS, 397, 1836,
  \dodoi{10.1111/j.1365-2966.2009.15123.x}

\bibitem[{Harris {et~al.}(2020)Harris, Millman, van~der Walt, Gommers,
  Virtanen, Cournapeau, Wieser, Taylor, Berg, Smith, Kern, Picus, Hoyer, van
  Kerkwijk, Brett, Haldane, del R{\'{i}}o, Wiebe, Peterson,
  G{\'{e}}rard-Marchant, Sheppard, Reddy, Weckesser, Abbasi, Gohlke, \&
  Oliphant}]{harris2020array}
Harris, C.~R., Millman, K.~J., van~der Walt, S.~J., {et~al.} 2020, Nature, 585,
  357, \dodoi{10.1038/s41586-020-2649-2}

\bibitem[{Heida {et~al.}(2019)Heida, Harrison, Brightman, F{\"u}rst, Stern, \&
  Walton}]{heidaSearchingDonorStars2019}
Heida, M., Harrison, F.~A., Brightman, M., {et~al.} 2019, The Astrophysical
  Journal, 871, 231, \dodoi{10.3847/1538-4357/aafa77}

\bibitem[{Hunter(2007)}]{Hunter:2007}
Hunter, J.~D. 2007, Computing in Science \& Engineering, 9, 90,
  \dodoi{10.1109/MCSE.2007.55}

\bibitem[{Huppenkothen {et~al.}(2016)Huppenkothen, Bachetti, Stevens, Migliari,
  \& Balm}]{huppenkothenStingraySpectraltimingSoftware2016}
Huppenkothen, D., Bachetti, M., Stevens, A.~L., Migliari, S., \& Balm, P. 2016,
  Astrophysics Source Code Library, ascl:1608.001

\bibitem[{Illarionov \& Sunyaev(1975)}]{illarionovWhyNumberGalactic1975}
Illarionov, A.~F., \& Sunyaev, R.~A. 1975, A\&A, 39, 185

\bibitem[{Israel {et~al.}(2017)Israel, Belfiore, Stella, Esposito, Casella,
  De~Luca, Marelli, Papitto, Perri, Puccetti, Castillo, Salvetti, Tiengo,
  Zampieri, D'Agostino, Greiner, Haberl, Novara, Salvaterra, Turolla, Watson,
  Wilms, \& Wolter}]{israelAccretingPulsarExtreme2017}
Israel, G.~L., Belfiore, A., Stella, L., {et~al.} 2017, Science, 355, 817,
  \dodoi{10.1126/science.aai8635}

\bibitem[{Jiang {et~al.}(2014)Jiang, Stone, \&
  Davis}]{jiangGlobalThreedimensionalRadiation2014}
Jiang, Y.-F., Stone, J.~M., \& Davis, S.~W. 2014, The Astrophysical Journal,
  796, 106, \dodoi{10.1088/0004-637X/796/2/106}

\bibitem[{Joss \& Rappaport(1984)}]{jossNeutronStarsInteracting1984}
Joss, P.~C., \& Rappaport, S.~A. 1984, Annual Review of Astronomy and
  Astrophysics, 22, 537, \dodoi{10.1146/annurev.aa.22.090184.002541}

\bibitem[{Kaaret {et~al.}(2017)Kaaret, Feng, \&
  Roberts}]{kaaretUltraluminousXRaySources2017}
Kaaret, P., Feng, H., \& Roberts, T.~P. 2017, Annual Review of Astronomy and
  Astrophysics, 55, 303, \dodoi{10.1146/annurev-astro-091916-055259}

\bibitem[{Kelley {et~al.}(1980)Kelley, Rappaport, \&
  Petre}]{kelleyDeterminationOrbitGX1980}
Kelley, R., Rappaport, S., \& Petre, R. 1980, The Astrophysical Journal, 238,
  699, \dodoi{10.1086/158026}

\bibitem[{King \& Lasota(2020)}]{kingPulsingNonpulsingULXs2020}
King, A., \& Lasota, J.-P. 2020, Monthly Notices of the Royal Astronomical
  Society, 494, 3611, \dodoi{10.1093/mnras/staa930}

\bibitem[{King {et~al.}(2017)King, Lasota, \&
  Klu{\'z}niak}]{kingPulsingULXsTip2017}
King, A., Lasota, J.-P., \& Klu{\'z}niak, W. 2017, MNRAS Let., 468, L59,
  \dodoi{10.1093/mnrasl/slx020}

\bibitem[{King(2008)}]{kingAccretionRatesBeaming2008}
King, A.~R. 2008, MNRAS Let., 385, L113,
  \dodoi{10.1111/j.1745-3933.2008.00444.x}

\bibitem[{Kosec {et~al.}(2018)Kosec, Pinto, Walton, Fabian, Bachetti,
  Brightman, F{\"u}rst, \& Grefenstette}]{kosecEvidenceVariableUltrafast2018}
Kosec, P., Pinto, C., Walton, D.~J., {et~al.} 2018, Monthly Notices of the
  Royal Astronomical Society, 479, 3978, \dodoi{10.1093/mnras/sty1626}

\bibitem[{Lam {et~al.}(2015)Lam, Pitrou, \& Seibert}]{numba2015}
Lam, S.~K., Pitrou, A., \& Seibert, S. 2015, in Proceedings of the Second
  Workshop on the LLVM Compiler Infrastructure in HPC, LLVM '15 (New York, NY,
  USA: Association for Computing Machinery), \dodoi{10.1145/2833157.2833162}

\bibitem[{Lange {et~al.}(2001)Lange, Camilo, Wex, Kramer, Backer, Lyne, \&
  Doroshenko}]{langePrecisionTimingMeasurements2001}
Lange, C., Camilo, F., Wex, N., {et~al.} 2001, MNRAS, 326, 274,
  \dodoi{10.1046/j.1365-8711.2001.04606.x}

\bibitem[{Levine {et~al.}(1993)Levine, Rappaport, Deeter, Boynton, \&
  Nagase}]{levineDiscoveryOrbitalDecay1993}
Levine, A., Rappaport, S., Deeter, J.~E., Boynton, P.~E., \& Nagase, F. 1993,
  ApJ, 410, 328, \dodoi{10.1086/172750}

\bibitem[{Levine {et~al.}(2000)Levine, Rappaport, \&
  Zojcheski}]{levineOrbitalDecayLMC2000}
Levine, A.~M., Rappaport, S.~A., \& Zojcheski, G. 2000, ApJ, 541, 194,
  \dodoi{10.1086/309398}

\bibitem[{{Lu} {et~al.}(2022){Lu}, {Fuller}, {Quataert}, \&
  {Bonnerot}}]{lu2022}
{Lu}, W., {Fuller}, J., {Quataert}, E., \& {Bonnerot}, C. 2022, arXiv e-prints,
  arXiv:2204.00847.
\newblock \doarXiv{2204.00847}

\bibitem[{Luo {et~al.}(2021)Luo, Ransom, Demorest, Ray, Archibald, Kerr,
  Jennings, Bachetti, {van Haasteren}, Champagne, Colen, Phillips, Zimmerman,
  Stovall, Lam, \& Jenet}]{luoPINTModernSoftware2021}
Luo, J., Ransom, S., Demorest, P., {et~al.} 2021, The Astrophysical Journal,
  911, 45, \dodoi{10.3847/1538-4357/abe62f}

\bibitem[{Madsen {et~al.}(2019)Madsen, Hickox, Bachetti, Stern, Gellert,
  Garc{\'i}a, Kara, Brandt, Krawczynski, Lohfink, Brenneman, Christensen,
  Middleton, Hornstrup, Matt, Jaodand, Lansbury, Ricci, Fuerst, Ballantyne,
  Walton, Fabian, Della Monica~Ferreira, Pottschmidt, Miller, Windt,
  Balokovi{\'c}, Kamraj, Wilms, Heida, Alexander, Boorman, Wik, Vogel,
  Earnshaw, Descalle, Civano, Fornasini, Grindlay, Zhang, Hornschemeier, \&
  Craig}]{madsenHEXPHighEnergyXray2019}
Madsen, K., Hickox, R., Bachetti, M., {et~al.} 2019, 51, 166

\bibitem[{Mellah {et~al.}(2019)Mellah, Sundqvist, \&
  Keppens}]{mellahWindRocheLobe2019}
Mellah, I.~E., Sundqvist, J.~O., \& Keppens, R. 2019, A\&A, 622, L3,
  \dodoi{10.1051/0004-6361/201834543}

\bibitem[{Middleton {et~al.}(2022)Middleton, Higginbottom, Knigge, Khan, \&
  Wiktorowicz}]{middletonThermallyDrivenWinds2022}
Middleton, M.~J., Higginbottom, N., Knigge, C., Khan, N., \& Wiktorowicz, G.
  2022, Monthly Notices of the Royal Astronomical Society, 509, 1119,
  \dodoi{10.1093/mnras/stab2991}

\bibitem[{Middleton {et~al.}(2021)Middleton, Walton, Alston, Dauser,
  Eikenberry, Jiang, Fabian, Fuerst, Brightman, Marshall, Parker, Pinto,
  Harrison, Bachetti, Altamirano, Bird, Perez, {Miller-Jones}, Charles, Boggs,
  Christensen, Craig, Forster, Grefenstette, Hailey, Madsen, Stern, \&
  Zhang}]{middletonNuSTARRevealsHidden2021}
Middleton, M.~J., Walton, D.~J., Alston, W., {et~al.} 2021, Monthly Notices of
  the Royal Astronomical Society, 506, 1045, \dodoi{10.1093/mnras/stab1280}

\bibitem[{Misra {et~al.}(2020)Misra, Fragos, Tauris, Zapartas, \&
  {Aguilera-Dena}}]{misraOriginPulsatingUltraluminous2020}
Misra, D., Fragos, T., Tauris, T.~M., Zapartas, E., \& {Aguilera-Dena}, D.~R.
  2020, Astronomy and Astrophysics, 642, A174,
  \dodoi{10.1051/0004-6361/202038070}

\bibitem[{Mushtukov {et~al.}(2017)Mushtukov, Suleimanov, Tsygankov, \&
  Ingram}]{mushtukovOpticallyThickEnvelopes2017}
Mushtukov, A.~A., Suleimanov, V.~F., Tsygankov, S.~S., \& Ingram, A. 2017,
  Monthly Notices of the Royal Astronomical Society, 467, 1202,
  \dodoi{10.1093/mnras/stx141}

\bibitem[{Mushtukov {et~al.}(2015)Mushtukov, Suleimanov, Tsygankov, \&
  Poutanen}]{mushtukovMaximumAccretionLuminosity2015}
Mushtukov, A.~A., Suleimanov, V.~F., Tsygankov, S.~S., \& Poutanen, J. 2015,
  MNRAS, 454, 2539, \dodoi{10.1093/mnras/stv2087}

\bibitem[{Pinto {et~al.}(2016)Pinto, Fabian, Middleton, \&
  Walton}]{pintoUltrafastOutflowsUltraluminous2016}
Pinto, C., Fabian, A., Middleton, M., \& Walton, D. 2016, arXiv,
  arXiv:1611.00623.
\newblock \doarXiv{1611.00623}

\bibitem[{Pletsch \& Clark(2015)}]{pletschGammaRayTimingRedback2015}
Pletsch, H.~J., \& Clark, C.~J. 2015, The Astrophysical Journal, 807, 18,
  \dodoi{10.1088/0004-637X/807/1/18}

\bibitem[{Poutanen {et~al.}(2007)Poutanen, Lipunova, Fabrika, Butkevich, \&
  Abolmasov}]{poutanenSupercriticallyAccretingStellar2007}
Poutanen, J., Lipunova, G., Fabrika, S., Butkevich, A.~G., \& Abolmasov, P.
  2007, MNRAS, 377, 1187, \dodoi{10.1111/j.1365-2966.2007.11668.x}

\bibitem[{Ray {et~al.}(2018)Ray, Arzoumanian, Brandt, Burns, Chakrabarty,
  Feroci, Gendreau, Gevin, Hernanz, Jenke, Kenyon, G{\'a}lvez, Maccarone,
  Okajima, Remillard, Schanne, Tenzer, Vacchi, {Wilson-Hodge}, Winter, Zane,
  Ballantyne, Bozzo, Brenneman, Cackett, De~Rosa, Goldstein, Hartmann,
  McDonald, Stevens, Tomsick, Watts, Wood, \&
  Zoghbi}]{raySTROBEXProbeclassMission2018}
Ray, P.~S., Arzoumanian, Z., Brandt, S., {et~al.} 2018, Space Telescopes and
  Instrumentation 2018: Ultraviolet to Gamma Ray, 10699, 1069919,
  \dodoi{10.1117/12.2312257}

\bibitem[{Rodr{\'i}guez~Castillo {et~al.}(2020)Rodr{\'i}guez~Castillo, Israel,
  Belfiore, Bernardini, Esposito, Pintore, De~Luca, Papitto, Stella, Tiengo,
  Zampieri, Bachetti, Brightman, Casella, D'Agostino, Dall'Osso, Earnshaw,
  F{\"u}rst, Haberl, Harrison, Mapelli, Marelli, Middleton, Pinto, Roberts,
  Salvaterra, Turolla, Walton, \&
  Wolter}]{rodriguezcastilloDiscoveryPulsarDay2020}
Rodr{\'i}guez~Castillo, G.~A., Israel, G.~L., Belfiore, A., {et~al.} 2020, The
  Astrophysical Journal, 895, 60, \dodoi{10.3847/1538-4357/ab8a44}

\bibitem[{Romanova {et~al.}(2004)Romanova, Ustyugova, Koldoba, \&
  Lovelace}]{romanovaPropellerRegimeDisk2004}
Romanova, M.~M., Ustyugova, G.~V., Koldoba, A.~V., \& Lovelace, R. V.~E. 2004,
  ApJ, 616, L151, \dodoi{10.1086/426586}

\bibitem[{Shakura \& Sunyaev(1973)}]{shakuraBlackHolesBinary1973}
Shakura, N.~I., \& Sunyaev, R.~A. 1973, A\&A, 24, 337

\bibitem[{Soberman {et~al.}(1997)Soberman, Phinney, \& {van den
  Heuvel}}]{sobermanStabilityCriteriaMass1997}
Soberman, G.~E., Phinney, E.~S., \& {van den Heuvel}, E. P.~J. 1997, Astronomy
  and Astrophysics, v.327, p.620-635, 327, 620

\bibitem[{Stoyanov \& Zamanov(2009)}]{2009AN....330..727S}
Stoyanov, K.~A., \& Zamanov, R.~K. 2009, Astronomische Nachrichten, 330, 727,
  \dodoi{10.1002/asna.200811224}

\bibitem[{{Surkova} \& {Svechnikov}(2004)}]{surkovacatalogue}
{Surkova}, L.~P., \& {Svechnikov}, M.~A. 2004, VizieR Online Data Catalog

\bibitem[{Tauris \& Savonije(2001)}]{2001nsbh.conf..337T}
Tauris, T.~M., \& Savonije, G.~J. 2001, Spin-{{Orbit Coupling}} in {{X}}-Ray
  {{Binaries}}, Vol. 567 ({eprint: arXiv:astro-ph/0001014}), 337

\bibitem[{Tauris \& {van den
  Heuvel}(2006)}]{taurisFormationEvolutionCompact2006}
Tauris, T.~M., \& {van den Heuvel}, E. P.~J. 2006, Compact stellar X-ray
  sources, 623

\bibitem[{Tsygankov {et~al.}(2016)Tsygankov, Mushtukov, Suleimanov, \&
  Poutanen}]{tsygankovPropellerEffectAction2016}
Tsygankov, S.~S., Mushtukov, A.~A., Suleimanov, V.~F., \& Poutanen, J. 2016,
  MNRAS, 457, 1101, \dodoi{10.1093/mnras/stw046}

\bibitem[{{van den Heuvel}(1994)}]{1994inbi.conf..263V}
{van den Heuvel}, E. P.~J. 1994, Interacting Binaries: Topics in Close Binary
  Evolution., 263--474

\bibitem[{Vasilopoulos {et~al.}(2021)Vasilopoulos, Koliopanos, Haberl, Treiber,
  Brightman, Earnshaw, \& G{\'u}rpide}]{vasilopoulosChandraProbesXRay2021}
Vasilopoulos, G., Koliopanos, F., Haberl, F., {et~al.} 2021, The Astrophysical
  Journal, 909, 50, \dodoi{10.3847/1538-4357/abda49}

\bibitem[{Virtanen {et~al.}(2020)Virtanen, Gommers, Oliphant, Haberland, Reddy,
  Cournapeau, Burovski, Peterson, Weckesser, Bright, {van der Walt}, Brett,
  Wilson, Millman, Mayorov, Nelson, Jones, Kern, Larson, Carey, Polat, Feng,
  Moore, {VanderPlas}, Laxalde, Perktold, Cimrman, Henriksen, Quintero, Harris,
  Archibald, Ribeiro, Pedregosa, {van Mulbregt}, \& {SciPy 1.0
  Contributors}}]{2020SciPy-NMeth}
Virtanen, P., Gommers, R., Oliphant, T.~E., {et~al.} 2020, Nature Methods, 17,
  261, \dodoi{10.1038/s41592-019-0686-2}

\bibitem[{Walton {et~al.}(2016)Walton, Tomsick, Madsen, Grinberg, Barret,
  Boggs, Christensen, Clavel, Craig, Fabian, Fuerst, Hailey, Harrison, Miller,
  Parker, Rahoui, Stern, Tao, Wilms, \& Zhang}]{waltonSoftStateCygnus2016}
Walton, D.~J., Tomsick, J.~A., Madsen, K.~K., {et~al.} 2016, The Astrophysical
  Journal, 826, 87, \dodoi{10.3847/0004-637X/826/1/87}

\bibitem[{Wang(1996)}]{wangLocationInnerRadius1996}
Wang, Y.-M. 1996, The Astrophysical Journal, 465, L111, \dodoi{10.1086/310150}

\bibitem[{Xu \& Li(2017)}]{xuMagneticFieldUltraluminous2017}
Xu, K., \& Li, X.-D. 2017, The Astrophysical Journal, 838, 98,
  \dodoi{10.3847/1538-4357/aa65d5}

\bibitem[{Zhang {et~al.}(2016)Zhang, Feroci, Santangelo, Dong, Feng, Lu,
  Nandra, Wang, Zhang, Bozzo, Brandt, De~Rosa, Gou, Hernanz, {van der Klis},
  Li, Liu, Orleanski, Pareschi, Pohl, Poutanen, Qu, Schanne, Stella, Uttley,
  Watts, Xu, Yu, in~'t Zand, Zane, Alvarez, Amati, Baldini, Bambi, Basso,
  Bhattacharyya, Bellazzini, Belloni, Bellutti, Bianchi, Brez, Bursa, Burwitz,
  {Budtz-Jorgensen}, Caiazzo, Campana, Cao, Casella, Chen, Chen, Chen, Chen,
  Chen, Chen, Civitani, Zelati, Cui, Cui, Dai, Del~Monte, De~Martino,
  Di~Cosimo, Diebold, Dovciak, Donnarumma, Doroshenko, Esposito, Evangelista,
  Favre, Friedrich, Fuschino, Galvez, Gao, Ge, Gevin, Goetz, Han, Heyl, Horak,
  Hu, Huang, Huang, Hudec, Huppenkothen, Israel, Ingram, Karas, Karelin, Jenke,
  Ji, Kennedy, Korpela, Kunneriath, Labanti, Li, Li, Li, Liang, Limousin, Lin,
  Ling, Liu, Liu, Liu, Lu, Lund, Lai, Luo, Luo, Ma, Mahmoodifar, Marisaldi,
  Martindale, Meidinger, Men, Michalska, Mignani, Minuti, Motta, Muleri,
  Neilsen, Orlandini, Pan, Patruno, Perinati, Picciotto, Piemonte, Pinchera,
  Rachevski, Rapisarda, Rea, Rossi, Rubini, Sala, Shu, Sgro, Shen, Soffitta,
  Song, Spandre, Stratta, Strohmayer, Sun, Svoboda, Tagliaferri, Tenzer, Tong,
  Taverna, Torok, Turolla, Vacchi, Wang, Wang, Walton, Wang, Wang, Wang, Wang,
  Weng, Wilms, Winter, Wu, Wu, Xiong, Xu, Xue, Yan, Yang, Yang, Yang, Yuan,
  Yuan, Yuan, Zampa, Zampa, Zdziarski, Zhang, Zhang, Zhang, Zhang, Zhang,
  Zhang, Zheng, Zhou, \& Zhou}]{zhangEXTPEnhancedXray2016}
Zhang, S.~N., Feroci, M., Santangelo, A., {et~al.} 2016, arXiv:1607.08823
  [astro-ph], 99051Q, \dodoi{10.1117/12.2232034}

\end{thebibliography}
\bibliographystyle{aasjournal}
\appendix


\section{Eccentricity and semi-major axis}\label{sec:ecc}
Thanks to the additional counts coming from the data reduction described in \secref{sec:reduction}, we could obtain a more stringent upper limit on the eccentricity and verify the past estimates of the semimajor axis.

We created a piecewise spindown solution for PINT (using the \texttt{PiecewiseSpindown} model) using all the best estimates of the frequency and the frequency derivative listed in Table~\ref{tab:eph}, that served as a baseline for the subsequent calculations.
We used HENphaseogram \citep{bachettiHENDRICSHighENergy2018} to obtain times of arrival (TOA) in 120 high signal-to-noise time intervals between MJD 56685.7 and 56722.
Then, we used \texttt{pintk} to look for features in the timing residuals reminiscent of an eccentricity.
The Roemer delay gives the delay of the signal from the pulsar during its binary motion.
In the limit of small eccentricity, this delay can be expressed as
\begin{equation}\label{eq:roemer}
    \Delta_R = X\left[\sin\Phi + \left(\frac{\kappa}{2}\sin 2\Phi + \frac{\eta}{2}\cos 2\Phi \right)\right]
\end{equation}
where $X=a\sin i/c$,  $\Phi=2\pi/\Porb(T-T_{asc})$, $\omega$ is the angle of periastron, $\kappa=e\sin\omega$ and $\eta=e\cos\omega$
Hence, eccentricity should produce sinusoidal residuals with $P=\Porb/2$ and amplitude $e a\sin i/c$ in the times of arrival corrected with a circular orbit.
These features are not correlated with any other orbital parameter of interest (which produce features at the orbital period), and can be investigated independently.
The Tempo2/PINT timing model ELL1  \citep{langePrecisionTimingMeasurements2001} implements this correction.
Using PINT, we fit the best-fit residual from the best circular model with ELL1 and found no significant features reminiscent of an eccentricity.
The new 3-sigma upper limit on eccentricity, using the 0.0005 error bars from this fit, is around 0.0015, half the value quoted by B14.

Using \eqref{eq:roemer}, it is also possible to compare the effect of an error on $a\sin i/c$ with that on $T_{asc}$.
Neglecting the eccentricity, we get that a given error on $\Delta_R$ can be written as
\begin{equation}\label{eq:roemererr}
\delta\Delta_R \approx \delta X \sin \Phi + \delta T_{asc} \frac{2\pi X}{\Porb} \cos{\Phi}
\end{equation}

For \Mtwo, $a\sin i/c=22.218(5)$. Therefore, we can neglect the error on $X$ whenever the error on $T_{asc}$ satisfies
\begin{equation}
    \delta T_{asc} \gg \frac{\delta X}{X} \frac{\Porb}{2\pi} \approx \mathrm{8\,s}
\end{equation}

This is always true in this Paper (see \tabref{tab:obs}).
Later observations are not able to constrain both $\tasc$ and $X$, and thawing $X$ in the fit artificially increases the error bars without leading to a more precise estimate: for short observations, it correlates with \tasc and $\nudot$ and the fit yields unreasonable values both for $X$ and the other parameters.

\section{Mass transfer}\label{sec:transfer}
By differentiating the formula for the orbital angular momentum and Kepler's third Law, it can be shown how the orbital separation and the orbital period change as a response to mass transfer or angular momentum changes (e.g. \citealt{taurisFormationEvolutionCompact2006}):
\begin{equation}\label{eq:pdotvsmdot}
    \frac{2}{3} \frac{\Porbdot}{\Porb} = \frac{\dot{a}}{a} = 2\frac{\dot{J}}{J} -2\frac{\Mdot_d}{M_d} -2\frac{\Mdot_p}{M_p} + \frac{\Mdot_d + \Mdot_p}{M_d + M_p} -2\dot{e}e
\end{equation}
where $J$ is the total angular momentum of the system, $M_{p}$ and $M_d$ are the masses of the pulsar and the donor, $a$ is the orbital separation, $e$ is the eccentricity, and dots denote time derivatives.
$\Mdot_d$ is negative and is $\Mdot_p$ positive, because the pulsar is accreting from the donor.
The eccentricity of \Mtwo is consistent with 0 (see Appendix~\ref{sec:ecc}), as expected from a Roche Lobe-overflowing system, so it is likely that the last term in the equation can be neglected.
But admitting that an undetected tiny eccentricity exists in the system, in order to have a negative orbital derivative there should be a \textit{positive} change, or an increase, of eccentricity, which is implausible given that these systems tend to circularize over time.

A number of phenomena causing changes in orbital angular momentum are discussed in the literature, such as gravitational wave (GW) emission (important in very compact systems such as some binary neutron stars), spin-orbit coupling (when the Roche-filling star's rotation is not synchronized with the orbit), magnetic braking (studied in low-mass X-ray binaries), and mass loss when the ejected mass has specific angular momentum.
Given the large donor mass and orbital distance, we do not expect GW emission or magnetic braking to be significant.
Moreover, even though they disagree on the exact mass transfer rate, different authors agree that the system is undergoing a strong mass transfer \citep{kingPulsingNonpulsingULXs2020,mushtukovOpticallyThickEnvelopes2017}.
Such a mass transfer rate is difficult to reconcile with mechanisms other than Roche-Lobe overflow (such as wind accretion or even wind Roche Lobe overflow, \citealt{mellahWindRocheLobe2019}), and the synchronization timescales are so small that we can also neglect spin-orbit coupling \citep{2009AN....330..727S,2001nsbh.conf..337T}.
This leaves us with mass transfer and or mass loss from a circumbinary disk (see below) as the only likely sources of angular momentum drain.

\textit{Conservative} mass transfer has no angular momentum or mass losses from the system (i.e., $\Mdot_p=-\Mdot_d$ and $\dot{J}=0$).
In this case, Equation~\ref{eq:pdotvsmdot} reduces to:
\begin{equation}\label{eq:pdotcons}
    \frac{\Porbdot}{\Porb} = 3\frac{\Mdot_d}{M_d}\left( \frac{M_d}{M_p} - 1\right)
\end{equation}
It is clear that, for $M_d/M_p>1$, the system responds to a mass transfer from the donor ($\Mdot_d<0$) by decreasing the orbital period, as observed.

The non-conservative mass transfer case (when mass is lost from the system in any form) implies a change of the total angular momentum and can be studied by dividing the angular momentum term into different terms.
Following the approach by \citet{1994inbi.conf..263V,sobermanStabilityCriteriaMass1997,taurisFormationEvolutionCompact2006},

\begin{equation}\label{eq:jorb}
    \frac{\dot{J}}{J} = \frac{\alpha + \beta r^2 + \delta\gamma(1 + r)^2}{1 + r} \frac{\Mdot_d}{M_d}
\end{equation}
and
\begin{equation}\label{eq:mdotp}
    \Mdot_p = -(1 - \alpha -\beta -\delta) \Mdot_d
\end{equation}
where $r=M_d/M_p$, $\alpha$ indicates the fraction of matter lost directly from the donor%
\footnote{Note that in other papers, e.g. \citep{jossNeutronStarsInteracting1984}, $\alpha$ indicates the specific angular momentum. This can create confusion when comparing the different approaches.}%
, $\beta$ the fraction lost from fast winds close to the accretor, and $\delta$ the fraction lost in a circumbinary disk of radius $a_r = \gamma^2 a$.

It is interesting to show where the three angular momentum losses lead when they dominate the orbital evolution, by developing Equation~\ref{eq:pdotvsmdot} with Equation~\ref{eq:jorb} and~\ref{eq:mdotp}.

For the loss from the donor ($\alpha=1$):
\begin{equation}\label{eq:alpha}
\frac{\Porbdot}{\Porb} = \frac{3}{2}\frac{\Mdot_d}{M_d}\left(\frac{-r}{1+r}\right)
\end{equation}
Therefore, an isotropic mass loss from the donor leads to an \textit{expansion} of the orbit.

For the loss from the accretor ($\beta=1$):
\begin{equation}\label{eq:beta}
\frac{\Porbdot}{\Porb} = \frac{3}{2}\frac{\Mdot_d}{M_d}\left(\frac{2r^2 - r - 2}{1+r}\right)
\end{equation}
implying that isotropic mass loss from the accretor (e.g., with disk winds) still leads to a \textit{contraction} of the orbit.
This is what is believed to happen at extreme mass transfer rates, where we expect strong radiation-driven winds to be launched inside the spherization radius \citep{shakuraBlackHolesBinary1973}.
In the limit $r\gg1$, this is equivalent to the conservative case.

Finally, for the circumbinary disk ($\delta=1$), we have
\begin{equation}\label{eq:delta}
\frac{\Porbdot}{\Porb} = \frac{3}{2}\frac{\Mdot_d}{M_d}\left(\frac{2\gamma(1 + r)^2 - 2 - r}{1+r}\right)
\end{equation}
which, for $\gamma\geq1$ (disk radius larger than orbital separation) and $r>1$, also produces a contraction of the orbit.

To summarize, the orbital decay we observe is compatible with the effect of mass transfer between a more massive donor and a neutron star (with or without mass loss from the accretor), or with angular momentum loss through an equatorial circumbinary disk, possibly launched by the second Lagrangian point L2 (\citealt{taurisFormationEvolutionCompact2006}.
Due to the observation that matter is indeed accreting onto the neutron star, and that we observe many ULXs in nearby galaxies which suggests that this accretion regime is not too short-lived, our analysis favors conservative (or mildly non-conservative) mass transfer from a intermediate/high mass star, with no high-angular momentum mass loss mechanisms.
Again, we stress that fast winds from the region around the compact object do not change the results considerably.
Moreover, as we show in Appendix~\ref{sec:magnetic}, the Spherization radius is likely in the proximity of the magnetospheric radius, changing these estimates by a relatively small amount.

\section{Important radii}\label{sec:radii}
Around an accreting neutron star, we can define two important radii (see \citealt{frankAccretionPowerAstrophysics2002} for a comprehensive treatment): the first, the \textit{corotation radius} $\rco$, is the radius at which the Keplerian angular velocity in the disk equals the angular velocity of the neutron star:
\begin{equation}\label{eq:rco}
    \rco = {\left(
    \frac{GM \pspin^2}{4\pi^2}
    \right)}^{\frac{1}{3}}.
\end{equation}
where $\pspin$ is the rotation period of the neutron star, $M$ its mass, and $G$ the Universal Gravitational constant.

The second is called the \textit{magnetospheric radius}, or inner radius, or truncation radius.
Within this radius, the accretion disk gets disrupted, and matter gets captured by the magnetic field lines and conveyed to the magnetic poles of the neutron star:
\begin{equation}\label{eq:rmag}
    \rin = \xi {\left(\frac{\mu^4}{GM\dot{M}^2}\right)}^{\frac{1}{7}}
\end{equation}
where $\mu$ is the magnetic dipole moment, $\Mdot$ the mass accretion rate, and $\xi\sim0.5$ encodes a number of effects like the accretion geometry (e.g. disk versus isotropic accretion) and the details of the interaction between the plasma and the different components of the magnetic field.

According to accretion theory, the relative position of $\rco$ and $\rin$ is what determines whether a neutron star will \textit{spin up} (accelerate its rotation) during accretion or spin \textit{down} (slow down its rotation).
The matter captured by the magnetic field of the neutron star at a given radius is orbiting with a given angular velocity, and will transfer angular momentum to the neutron star through the magnetic field lines.
Outside $\rco$, this velocity is lower than the angular velocity of the neutron star, while it is higher inside.
Therefore, roughly speaking, if $\rin<\rco$ the star spins up, and if $\rin>\rco$ it spins down.
Various corrections can be made, integrating the torque from the matter outside and inside the corotation radius, and different authors come up with different prescriptions that can in general be treated by multiplying $\rin$ by a factor of order 1 \citep{ghoshDiskAccretionMagnetic1978,wangLocationInnerRadius1996}.
When $\rin\sim\rco$, small changes of accretion rate move the inner radius back and forth around $\rco$, and we can expect the source to alternatively spin up and down.
This situation is called \textbf{spin equilibrium}.

When $\rin\gg\rco$, the rotating magnetic field is able to swipe away the disk, and it is expected that accretion onto the neutron star will stop.
This is known as \textit{propeller} regime \citep{illarionovWhyNumberGalactic1975}.

Around a super-Eddington accreting source, a third important radius is often cited, the \textit{spherization} radius at which the disk departs from an ideal thin disk. Inside this radius, the mass in excess of the local Eddington limit is ejected in winds \citep{shakuraBlackHolesBinary1973}:
\begin{equation}
    \rsph = \frac{27}{4}R_g \frac{\Mdot}{\Mdot_{\rm Edd}}
\end{equation}
where the Eddington mass accretion rate $\Mdot_{\rm Edd}\approx \ledd/\eta c^2\approx 1.6\cdot10^{18}$\,g~s$^{-1}$ for a 1.4-$\msun$ neutron star, where $\eta\approx0.15$ is the efficiency, $R_g=GM/c^2$ is the gravitational radius, and $c$ is the speed of light.

\section{Donor star}\label{sec:donor}

The mass function determined through timing gives important insights on the kind of donor star we can expect:
\begin{equation}
    f = \frac{M^3_d \sin^3 i}{(M_p + M_d)^2} = \frac{\Omega^2_{\rm orb}}{G} (a_p\sin i)^3\approx1.83\,\msun
\end{equation}
where $\Omega_{\rm orb}=2\pi/P_{\rm orb}$ is the orbital angular velocity, $a_p\sin i$ is the projected semi-major axis of the pulsar orbit, $M_d$ is the mass of the donor, $M_p$ is the mass of the pulsar, and $i$ is the inclination.
In the formula above, the $\Omega_{\rm orb}$ and $a_p \sin i$ are measured from pulsar timing, while the left-hand side can be used to infer the donor mass given reasonable assumptions about the pulsar mass and the inclination.

Since $\sin i$ cannot be larger than 1 (orbit edge-on), this poses a hard lower limit to the donor star mass, that cannot be less than $3.56\,\msun$ (assuming a neutron star mass of $1.4\,\msun$).
The absence of eclipses from a (most likely) Roche-Lobe filling donor pushes the lower limit to $\sim5\msun$ (B14) and corresponds to an upper limit on the inclination of $\sim60^{\circ}$.
An unlikely donor mass of $100\,\msun$ corresponds instead to an inclination of $\sim 17^{\circ}$, which we take as a lower limit.

Similar arguments can be used to constrain the donor radius.
Assuming Roche Lobe overflow, the size of the donor is fixed by the mass ratio and orbital separation.

\begin{figure}[ht]
    \centering
    \includegraphics[width=\linewidth]{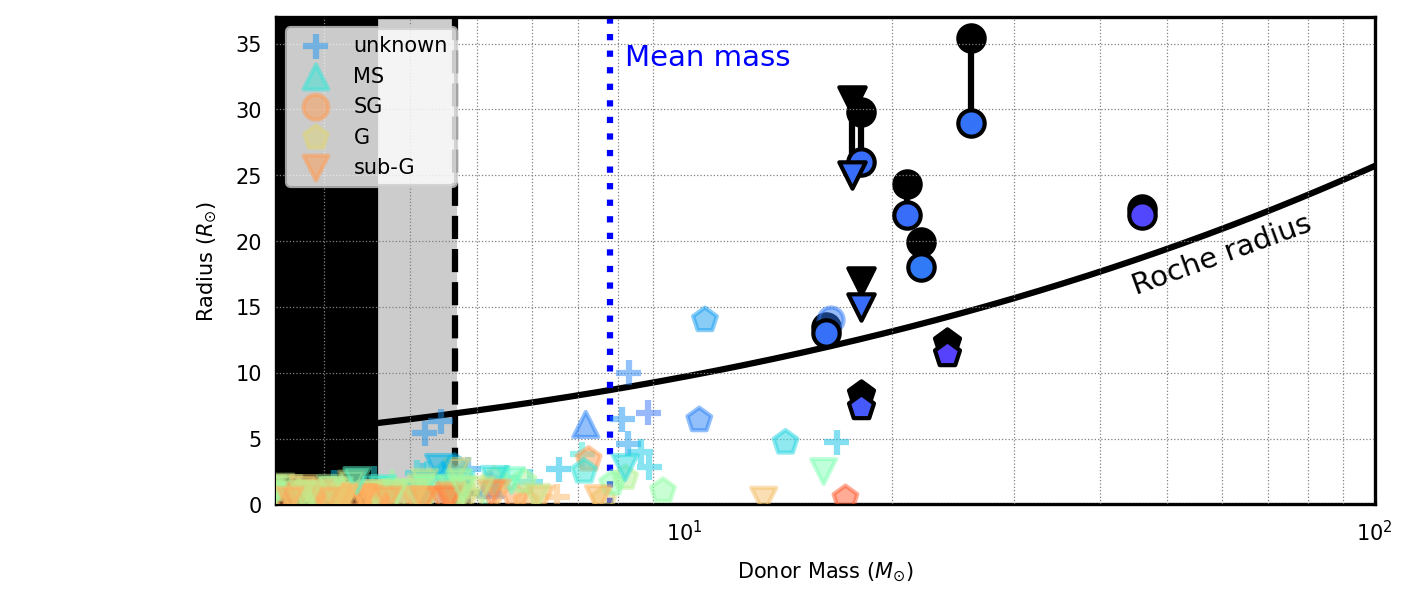}
    \caption{Roche Lobe radius in Eggleton approximation \citep{eggletonApproximationsRadiiRoche1983} versus mass for the donor star in \Mtwo.
    The donor has to lie around the black solid line in order to undergo Roche Lobe overflow.
    We overplot all donors from the HMXBs in \citet{falangaEphemerisOrbitalDecay2015} and all semidetached binary stars from \citet{surkovacatalogue} for comparison.
    For the HMXBs, we plot in black the Roche Lobe radius.
    Colors span the A (red)--O (blue) spectral types, and markers indicate different branches in the HR diagram.
    The grey shaded area is excluded by the absence of eclipses.
    The black area is prohibited by the mass function and the necessity that $\sin i \leq 1$.}
    \label{fig:radiusmass}
\end{figure}

With these constraints in mind (see \figref{fig:radiusmass}), and compared with known populations of donor stars in HMXBs, the most probable candidates are O/B giant stars between 5--100$\msun$.
Between $\sim17^{\circ}$ ($100\,\msun$ donor) and $\sim60^{\circ}$ ($5\,\msun$ donor), we assume all orientations to be equally probable.
This means that the values of the cosine of the inclination are equally probable between the two limiting cases $\cos 60^{\circ}$ and $\cos 17^{\circ}$.
This gives an average inclination of $\sim43^{\circ}$, corresponding to a donor mass of $\sim7.8\,\msun$.
Note that an archival search in HST data found several stars of this range of masses which could in principle be the donor \citep{heidaSearchingDonorStars2019}.

\section{Magnetic field estimates}\label{sec:magnetic}
\begin{figure}[ht]
    \centering
    \includegraphics[width=\linewidth]{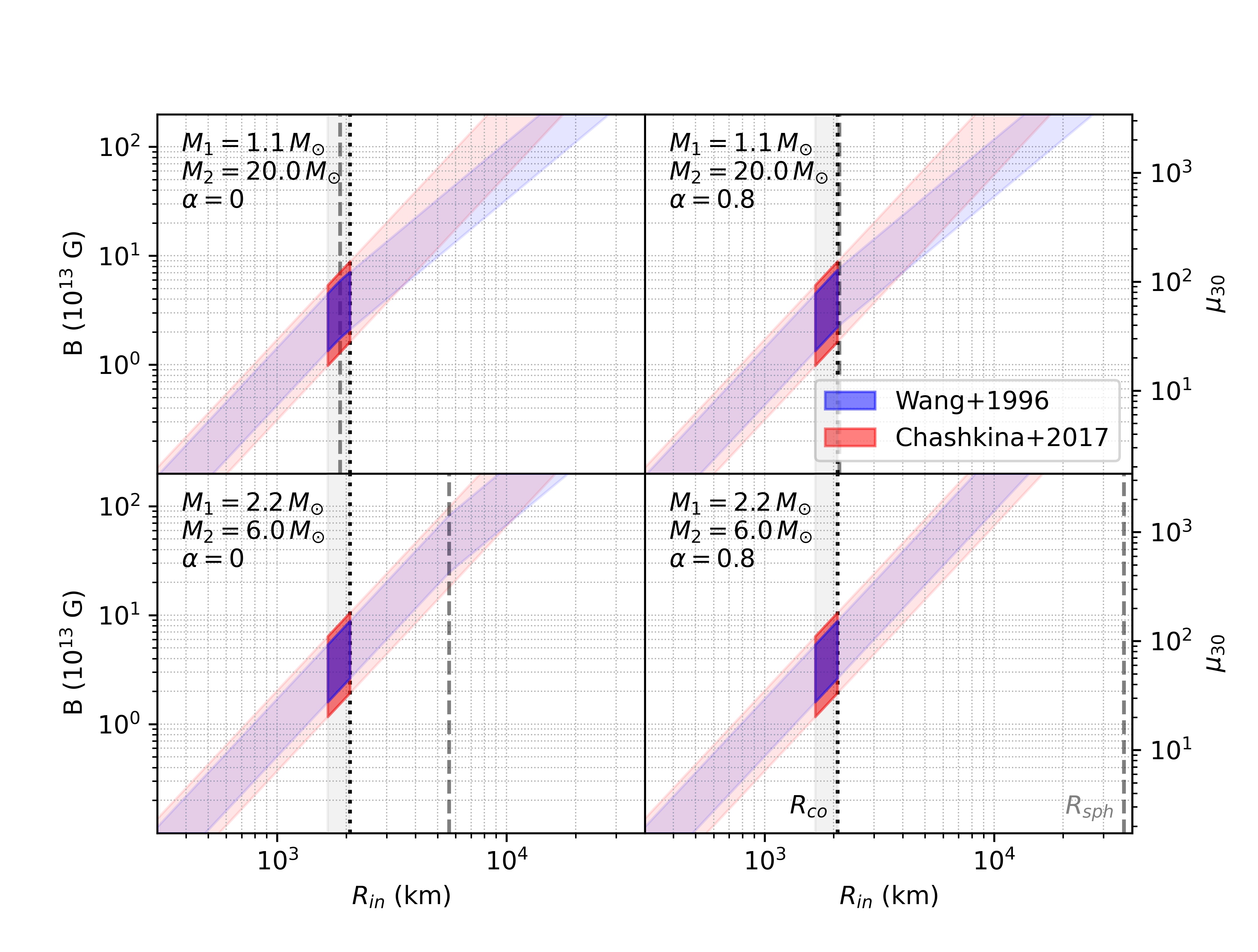}
    \caption{neutron star dipolar magnetic field estimate assuming spin equilibrium as described in the text, comparing the models from \citet{wangLocationInnerRadius1996} (the area corresponds to values of $0.5<\xi<1$) and from \citet{chashkina2017} (the area covers the range of viscosity parameter $0.01<\alpha_v<0.3$), in four cases with a range of mass ratios and a different mass loss fraction ($\alpha$ in Equation~\ref{eq:jorb}) from the donor.
    We show $\rco$ and $\rsph$ with vertical lines.
    Note that a lower mass ratio increases the estimated mass transfer. Also, the mass loss from the donor implies a larger mass transfer rate to justify the observed derivative. This explains the larger $\rsph$. We highlight the region where $0.8 <\rin/\rco < 1$, as expected from spin equilibrium.
    For the traditional model we took into account a change of accretion rate due to mass loss inside $\rsph$.
    The estimated dipolar magnetic field is always above $10^{13}$\,G for reasonable values of the parameters.}
    \label{fig:bfield}
\end{figure}

Traditional models, such as those proposed by \citet{ghoshDiskAccretionMagnetic1978} or \citet{wangLocationInnerRadius1996}, consider a thin disk with negligible radiation effects, and the inner radius is given by Equation~\ref{eq:rmag}.
Therefore, given a mass accretion rate, the position of the inner radius in this model is a function of the dipolar component of the magnetic field, modulo the order-unity constant $\xi$.
Since the source is close to spin equilibrium, as demonstrated by the spin behavior over time (see \tabref{tab:obs}), the inner radius has to be close to the corotation radius.
Therefore, equating the inner radius to the corotation radius, we can get an estimate of the magnetic field, as shown in \figref{fig:bfield} with the blue band.
Here, we take into account mass losses in a wind inside $\rsph$, when relevant, until $\rin$.
We use the "classical" mass loss obtained when the effects of advection are neglected \citep{shakuraBlackHolesBinary1973}. In this case, the accretion rate drops linearly with radius, and thus an upper limit on the accretion rate (i.e., an upper limit on the mass loss rate) corresponds to an upper limit on $\rin$ and a lower limit to the magnetic field strength.
Despite this conservative approach, the estimate on the magnetic field is robustly above $10^{13}$\,G.
Most models for sub-Eddington accretion agree within an order of magnitude for the treatment of spin equilibrium (see, e.g., \citealt{chenStudyingMagneticFields2021}).
However, it is possible that these models which are based on the interaction of a magnetized neutron star with a thin disk, with no radiation pressure either from the disk or from the central object, need to be corrected in the case of super-Eddington disks.
\citet{chashkina2017,chashkinaSuperEddingtonAccretionDiscs2019} have investigated this issue, finding that, indeed, the disk structure changes significantly when radiation pressure becomes dominant.
In particular, they find that $\xi$ is not constant, but depends on local (inside the disk) and external (e.g. from the neutron star) radiation pressure, and the amount of advection in the disk.
With the transfer rate $>100\times$Eddington we infer in this Paper, the inner radius becomes almost independent of the mass accretion rate and is described by Eq. 61 from \citet{chashkina2017}:
\begin{equation}\label{eq:chashkina}
    \frac{\rin}{R_{g}} \approx {\left(\frac{73\alpha}{24}\right)}^{2/9}{\left[\lambda{\left(\frac{\mu}{10^{30}\mathrm{G\,cm^3}}\right)}^2\right]}^{2/9}
\end{equation}
where $\alpha\sim0.1$ is the viscosity in the disk and $\lambda\sim4\cdot10^{10}\,(M_p/1.4\msun)^{-5}$.
\figref{fig:bfield} shows that, for a reasonable range of the viscosity parameter
\footnote{We call this parameter $\alpha_v$ instead of $\alpha$ to avoid confusion with the mass loss parameter $\alpha$}
$0.01<\alpha_v<0.3$, the estimate of the magnetic field obtained by equating the inner radius to the corotation radius using Equation~\ref{eq:chashkina} is similar to the prediction of traditional models using Equation~\ref{eq:rmag}, confirming an estimated magnetic field for \Mtwo\ above $10^{13}$\,G, as estimated with the classical model and by other authors in the literature \citep{tsygankovPropellerEffectAction2016,chenStudyingMagneticFields2021}.

\section{Pulsed fraction in the XMM-Newton energy band}\label{sec:energy}

\begin{figure}
    \centering
    \includegraphics[width=\linewidth]{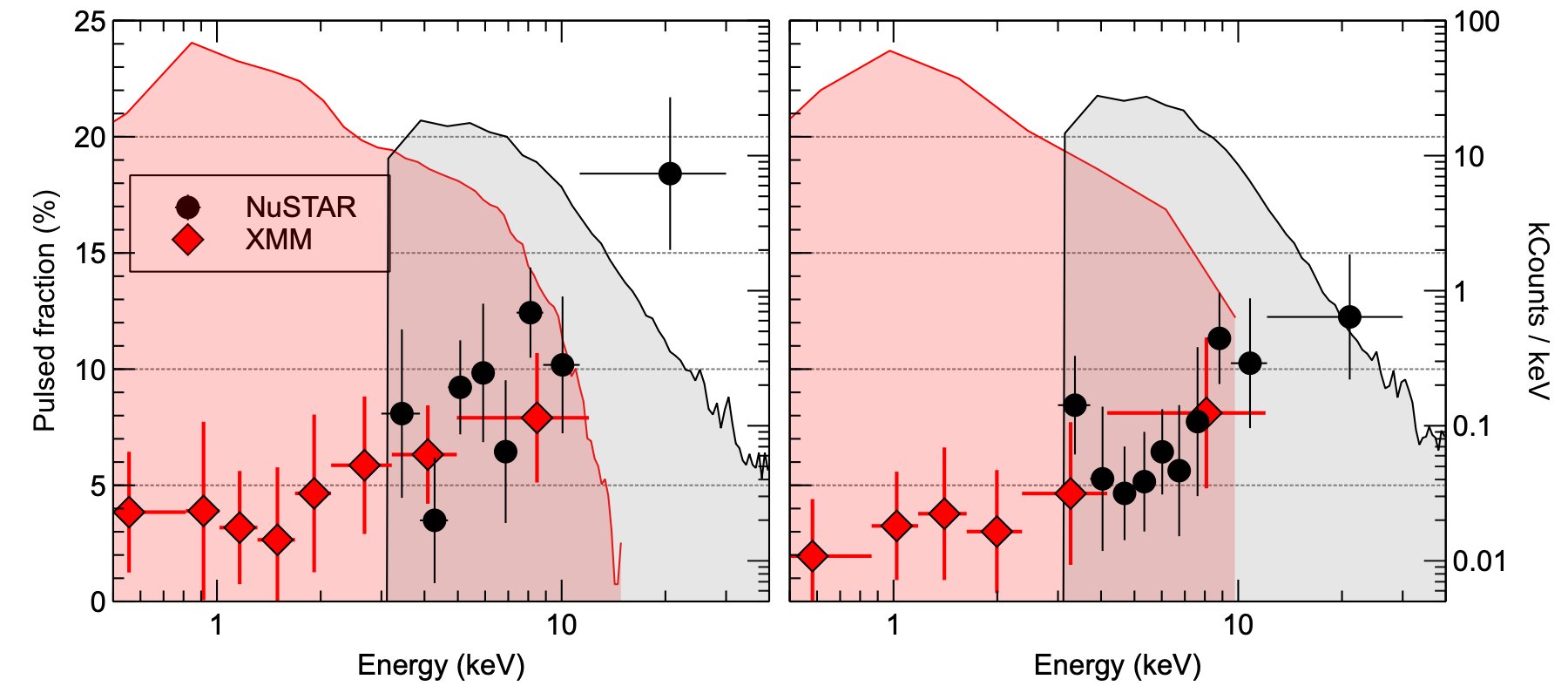}
    \caption{Pulsed amplitude versus energy, using \xmm data from obsIDs 0870940101 and 0870940401, and \nustar data from obsIDs 30602027002 and 30602027004.
    We overplot the raw count spectra of NuSTAR and XMM, showing how \xmm has many more counts, but in energy intervals with little or no pulsations.}
    \label{fig:xmm_pulse_amp}
\end{figure}

As opposed to many other PULXs, \Mtwo is very difficult to study with \xmm. Pulsations were not detected in many past observations of \Mtwo, despite the higher angular resolution of the EPIC-pn instrument.
One of the reasons is the lack of pulsations below 3 keV, due to both an intrinsic low pulsed amplitude and the very strong emission of \Mone and the M82 galaxy itself that increase the background at low energies.
In the 2021 quasi-simultaneous observations with \xmm and \nustar, we did manage to detect pulsations with EPIC-pn (\figref{fig:xmm_pulse_amp}).
\Mtwo was observed by \xmm on UT 2021-04-06 and 2021-04-16 for a total on-source exposure time of $\sim$70\,ks.
The only camera onboard \xmm that is able to detect pulsations from \Mtwo is EPIC-pn, that was set in Full Window mode.

We downloaded the data from the two observations from the \xmm archive\footnote{nxsa.esac.esa.int}
and processed them with the Science analysis software (SAS) version 20211130.

We ran the standard pipeline, using the tool \texttt{epchain} to obtain cleaned event files.
The M82 field is very crowded, and it is not possible to separate the emission of \Mtwo, \Mone and the diffuse Galactic center emission.
However, being mostly interested in the timing properties of \Mtwo, a precise modeling of the background is not strictly needed.
We selected photons coming from a region of $\sim50''$ around the putative position of \Mtwo
We cleaned the data from periods of high background activity.
Finally, we barycentered the data using the tool \texttt{barycen} using the \chandra position of \Mtwo, with the same ephemeris used in \texttt{barycorr}.

After this pre-processing, we folded the cleaned and barycentered event lists at the ephemeris obtained from the nearest \nustar observations, slightly adjusting the spin frequency through the maximization of the Rayleigh test.
We calculated the pulsed fraction from a sinusoidal modeling of the pulsed profile, as (Max - Min) / (Max + Min).
We plot this pulsed fraction, and the corresponding pulsed fraction from the quasi-simultaneous \nustar observations, in \figref{fig:xmm_pulse_amp}

\end{document}